\documentclass{bmcart}

\usepackage{amsmath,amssymb,amsfonts}
\usepackage{graphicx}
\usepackage{xcolor,color,colortbl}
\usepackage{algorithm}
\usepackage{algorithmic}
\usepackage{changes}
\usepackage{enumitem}
\usepackage{epstopdf}
\usepackage[numbers,sort&compress]{natbib} 
\newcommand{\bigO}{\mathcal{O}}
\DeclareMathOperator*{\argmin}{argmin}

\newcommand*{\argminl}{\argmin\limits}

\startlocaldefs
\endlocaldefs

\begin{document}
\begin{frontmatter}
\begin{fmbox}
\dochead{Research}

\title{Primary User Emulation and Jamming Attack Detection in Cognitive Radio via Sparse Coding}

\author[
   addressref={aff1},                   % id's of addresses, e.g. {aff1,aff2}
   corref={aff1}, 
   noteref={n1},                        % id's of article notes, if any
   email={haji.madni@std.medipol.edu.tr}   % email address
]{\inits{HM}\fnm{Haji M} \snm{Furqan}}
\author[
   addressref={aff1},                   % id's of addresses, e.g. {aff1,aff2}
   noteref={n1},                        % id's of article notes, if any
   email={mehmet.aygul@std.medipol.edu.tr}   % email address
]{\inits{MA}\fnm{Mehmet A} \snm{Ayg\"{u}l}}
\author[
   addressref={aff1},                   % id's of addresses, e.g. {aff1,aff2}
   email={mahmoud.nazzal@emu.edu.tr}   % email address
]{\inits{M}\fnm{Mahmoud} \snm{Nazzal}}
\author[
   addressref={aff1,aff2},
   email={huseyinarslan@medipol.edu.tr}
]{\inits{H}\fnm{H\"{u}seyin} \snm{Arslan}}

\address[id=aff1]{%                           % unique id
  \orgname{Department of Electrical and Electronics Engineering, Istanbul Medipol University, Istanbul, 34810 Turkey}, % university, etc
  %\street{Waterloo Road},                     %
  \postcode{34810}                                % post or zip code
  \city{Istanbul},                              % city
  \cny{TR}                                    % country
}
\address[id=aff2]{%                           % unique id
  \orgname{Department of Electrical Engineering, University of South Florida}, % university, etc
  %\street{Waterloo Road},                     %
  \postcode{3620}                                % post or zip code
  \city{Tampa},                              % city
  \cny{USA}                                    % country
}

\begin{artnotes}
\note[id=n1]{Equal contributor} % note, connected to author
\end{artnotes}

\end{fmbox}

\begin{abstractbox}
\begin{abstract} 
Cognitive radio is an intelligent and adaptive radio that improves the utilization of the spectrum by its opportunistic sharing. However, it is inherently vulnerable to primary user emulation and jamming attacks that degrade the spectrum utilization. In this paper, an algorithm for the detection of primary user emulation and jamming attacks in cognitive radio is proposed. The proposed algorithm is based on the sparse coding of the compressed received signal over a channel-dependent dictionary. More specifically, the convergence patterns in sparse coding according to such a dictionary are used to distinguish between a spectrum hole, a legitimate primary user, and an emulator or a jammer. The process of decision-making is carried out as a machine learning-based classification operation. Extensive numerical experiments show the effectiveness of the proposed algorithm in detecting the aforementioned attacks with high success rates. This is validated in terms of the confusion matrix quality metric. Besides, the proposed algorithm is shown to be superior to energy detection-based machine learning techniques in terms of receiver operating characteristics curves and the areas under these curves.
\end{abstract}
\end{abstractbox}
\end{frontmatter}
\section{Introduction} 
\label{Section1}

\par Due to the rapid growth in wireless technology and services, the scarcity of the wireless spectrum has become a major problem \cite{arjoune2019}. To meet the requirements of future wireless networks and to alleviate this spectrum scarcity problem, cognitive radio (CR) is one of the most promising solutions. CR allows spectrum sharing between the primary users (PUs) and secondary users (SUs). More specifically, it enables SUs to opportunistically utilize empty spectrum bands without harming the PUs by following these steps: i) determining whether the channel is occupied or not, ii) choosing the best part of the spectrum based on their quality of service (QoS) requirements, iii) coordinating with other users to access the spectrum, and iv) leaving the channel whenever a PU starts to transmit its data \cite{Yucek}.

\par Although CR is a promising solution to address the spectrum shortage problem, it is inherently vulnerable to both traditional and new security threats \cite{securcong}. This is due to the wireless nature and unique characteristics of CR. Traditional security threats include eavesdropping, spoofing, and jamming attacks \cite{jm5738229}, while new security threats include spectrum sensing data falsification (SSDF) and primary user emulation attack (PUEA) \cite{securcong, BOUABDELLAH201840}.

\par An eavesdropper tries to ``hear" the secret communication between legitimate nodes while a spoofer can modify, intercept, and replace the messages between the legitimate parties. On the other hand, a jammer can generate intentional interference signals to degrade the quality of communication for both PUs and SUs. Thus, a jammer can also prevent an SU from efficiently utilizing the white spaces of the spectrum by causing false alarms regarding spectrum occupancy \cite{jm5738229}.

\par In an SSDF attack, an illegitimate node provides false sensing information to degrade the performance of the collaborative spectrum sensing approach, where collaborative approaches include the interaction of multiple CRs to improve the sensing performance in the fading environment. On the other hand, PUEA is based on emulating the characteristics of the PU transmission to deceive the SUs about spectrum occupancy. PUEA prevents them from utilizing the existing spectrum holes and can even cause interference to the PUs in some cases \cite{PUEAnSSDF}.

\par  Popular PUEA types include malicious and selfish attacks. The malicious attackers’ objective is to degrade the CRs performance by preventing them from opportunistic exploitation of spectrum. Particularly, a malicious attacker destroys the operations of the CR network. Thus, it can stop CRs from sensing and can also disengage the already used spectrum by them. On the other hand, a selfish attacker aims at exploiting the space of the spectrum by preventing other secondary users from using it. More specifically, it focuses on enhancing its consumption of the spectrum by degrading the overall fairness of the system.

\par The focus of this work is to detect false alarm about the spectrum occupancy that is caused by illegitimate nodes. An illegitimate node can transmit a signal similar to that of a PU, considered as a primary user emulator (PUE), or can send an unstructured signal, considered as a jamming attack. In the literature, several solutions are proposed for illegitimate node detection. For instance, the power level of the signal through the energy detection (ED) algorithm can decide on the source of the signal \cite{7366825} using a pre-defined threshold. In \cite{6331677}, the authors presented a Markov random field-based belief propagation framework with ED for PUEA detection. Firstly, SUs employ the energy-based algorithm and calculate the belief values about the real source of the signal. Afterwards, the belief values are shared between different users. Finally, the average belief value is compared with the pre-defined threshold. If the average is less than the threshold, it is assumed that the signal source is fake, otherwise, the source of the signal is assumed to be real. These approaches are simple, however, they are shown to create high levels of false alarm rates. Cross-layer techniques are also effective for illegitimate node detection. In \cite{6504232}, the authors proposed a cross-layer approach for jamming attack and PUEA detection in CR networks by using information from physical layer spectrum sensing, statistical analysis of routing information, and prior knowledge about PUs. This technique is effective for detecting PUEA and jamming attack. However, there is an excessive overhead in analyzing and comparing information from physical and network layers.

\par The wireless channel and inherent physical characteristics of communication devices are also effective for illegitimate node detection \cite{rehman2014radio, 6239877,li2019detecting,6096448}. For instance, wireless channel-based detection schemes are proposed in \cite{rehman2014radio, 6239877, li2019detecting} for PUEA detection. These techniques are based on the fact that the channel between different transmitter-receiver pairs is different due to its spatial decorrelation nature. In \cite{6096448}, the inherent physical layer features of devices based on hardware impairments are exploited for PUEA detection. Nevertheless, these techniques require excessive software and hardware overheads for their implementation.

\par Localization-based detection is also popular for PUEA detection. The basic idea is to infer the position of the signal's source by using the received signal and compare it with a database of pre-known locations of legitimate PUs. However, database management is not applicable in all scenarios \cite{4413138,123loc}. Similarly, the authors in \cite{7450851} used the time difference of arrival-based position estimation approach for PUEA detection. However, this requires a strict synchronization between the receiver and the transmitter.
 
\par Machine learning (ML)-based solutions also received considerable attention for CR security. In \cite{liu2009aldo}, an anomaly detection framework for CR networks based on the characteristics of radio propagation is proposed. However, it does not consider specific attacks and is designed only for the detection of general anomalies. In \cite{luo2011specific}, the authors proposed a technique based on support vector data description (SVDD) and zoom fast Fourier transform (zoom FFT). In the first step, the pilot and symbol rate are estimated using zoom FFT. Afterwards, a boundary around the PU objects is constructed using the SVDD classifier which is used to distinguish between PU and PUE. However, this method does not perform well in low signal-to-noise ratio (SNR) operating conditions. Furthermore, the method fails when the PUE is extremely intelligent (the only information unknown by the PUE is the channel). In \cite{arul9svm}, the authors proposed an ML-based algorithm for PUEA detection that exploits the signal strength and boundaries around the position of PU for the correct detection. This method is good in terms of complexity but it suffers from performance degradation. 

\par Recently, compressive sensing (CS)-based approaches were applied in spectrum sensing where CS offers several benefits. For example, it can alleviate the need for high sampling rate analog-to-digital converters \cite{8052498, choi2017compressed, 8688400}. This results in a reduction of the overall complexity, energy consumption, and memory requirements. Following its success in various application areas \cite{8052498}, CS has been applied to the problem of PUEA detection. Works along this line include PUEA detection based on CS and received signal strength \cite{a9020025}. This approach needs multiple sensors throughout the network. Hence, it increases the overall complexity. Another example considers exploiting belief propagation and CS for PUEA detection \cite{7967058}. However, this requires a centralized node for its implementation. In \cite{7418201}, the authors proposed an algorithm for jamming attack detection in wide-band CR. In the first step, CS is performed to estimate a wide-band spectrum. Afterwards, an ED algorithm is applied to identify the occupied spectrum sub-bands. Lastly, waveform parameters of the sub-bands are compared with the known user database to determine the jamming attack. However, this method also requires database management. 

\par In this paper, we propose an algorithm for PUEA and jamming attack detection corresponding to the narrow-band spectrum using the convergence patterns of the sparse coding over channel-dependent sampled dictionary. This convergence is characterized by the sparse coding residual signal energy decay rates. The proposed algorithm does not require a centralized node or strict synchronization between transceiver ends. Moreover, it does not require information from multiple sensors for the implementation. Furthermore, it eliminates the need for estimating the sparse coding error tolerance or the sparsity level, as typically required in CS-based approaches. The reason is that the sparse recovery in the proposed algorithm is just used for energy convergence rate revelation rather than accurate signal reconstruction. The main contributions of this paper are as follows:

\begin{itemize}[leftmargin=*]

\item First, the decaying pattern of sparse coding is used for PUEA detection. This is achieved by exploiting the convergence patterns of the sparse coding over a PU channel-dependent dictionary. In this context, these patterns guide on identifying a spectrum hole, a PU, and a PUE through ML approaches.

\item Second, jamming attack detection is also performed based on the decay pattern of sparse coding. Here, the idea is that the noise and jamming signals are not compressible because they are not structured. So, residual energy decay patterns with a channel-dependent dictionary along with the non-compressive nature of jamming signals are used for efficient jamming attack detection via ML classification.
\end{itemize}

\par The rest of this paper is organized as follows. Preliminary information and the system model are presented in Section~\ref{Section2}. Section~\ref{Section3} provides the proposed algorithm, while the complexity analysis is presented in Section~\ref{Section4}. Section~\ref{Section5} presents the simulation results and discussions. Finally, the paper is concluded in Section~\ref{Section6}.

\par \textit{Notation}: Upper-case bold-faced, lower-case bold-faced and lower-case plain letters represent matrices, vectors, and scalars, respectively. The symbols $\|.\|_0$ and ${\|.\|}_2$ denote the number of nonzero elements and the 2-norm of a vector, respectively. The $\langle\cdot,\cdot\rangle$, $\dag$, and $\mathbb{C}$  symbols represent inner product, Moore-Penrose pseudoinverse, and complex number field.
\section{Preliminaries and System Model} \label{Section2}
This section reviews background information related to CS, sparse recovery, and ML approaches.

\subsection{Compressive Sensing and Sparse Recovery}

\par Using a random sensing matrix, CS merges data measurement and compression into a unified operation. CS applies to compressible signals, i.e., either the explicitly sparse signals, or the ones admitting sparsity in a certain domain \cite{davenport2010signal}. 

\par Let us assume a signal vector $\boldsymbol{y} \in \mathbb{C}^N$. A compressed version of $\boldsymbol{y}$ can be obtained by applying a measurement matrix $\boldsymbol{\Phi} \in \mathbb{C}^{M\times N}$ as $\boldsymbol{y_c}=\boldsymbol{\Phi}\boldsymbol{y}$, where $M \ll N$. Hence, a reduction in dimensionality from $N$-to-$M$ is achieved. A high-dimensional version of the original signal can be reconstructed from this low dimensional measurement via sparse recovery \cite{davenport2010signal}.

\par Generally speaking, let us assume that a signal $\boldsymbol{y}$ admits sparse coding over a dictionary ($\boldsymbol{D} \in \mathbb{C}^{N \times K}$). The signal can be represented in terms of $\boldsymbol{D}$ as $\boldsymbol{y}\approx\boldsymbol{Dw}$, where $\boldsymbol{w}\in\mathbb{C}^K$ is a sparse coefficient vector. The calculation of $\boldsymbol{w}$ can be cast as follows.
\begin{equation}
\argminl_{\boldsymbol{w}}{\|\boldsymbol{y}-\boldsymbol{Dw}\|}_2^2 ~ s.t. ~ {\|\boldsymbol{w}\|}_0 <S,\label{equa2}
\end{equation}
\noindent where $S$ denotes the sparsity level of the signal. Sparse recovery is an NP-hard problem. However, sparse recovery methods offer efficient approximate solutions. As shown in (\ref{equa2}), the $\ell_{0}$ pseudo-norm is principally used to exactly quantify the sparsity level. However, its minimization is mathematically intractable and highly complex. Therefore, there exist only approximate solutions to $\ell_{0}$ minimization, such as the matching pursuit and orthogonal matching pursuit (OMP) approaches. Alternatively, this problem can be overcome by relaxing the $\ell_{0}$ norm minimization condition to minimizing the $\ell_{1}$ norm which is a loose bound on sparsity. Still, $\ell_{1}$ minimization is convex and accepts linear programming. Thus, replacing $\ell_{0}$ minimization with $\ell_{1}$ minimization offers a significant reduction to the computational complexity of sparse coding. However, $\ell_{1}$ minimization requires information about the noise level of the signal being recovered. Thus, in this work, we adopt approximate $\ell_{0}$ minimization through the OMP algorithm \footnote{The proposed algorithm is not limited to OMP and it can be implemented with any sparse recovery algorithm \cite{8577023p}. We prefer to use the OMP algorithm since it is computationally efficient and simple.}.

\par The intrinsic sparsity of the signal can be revealed by a dictionary. This dictionary can be formed of fixed basis functions such as Fourier basis, Gabor functions, wavelets, and contourlets. Alternatively, it can be generated as a learned dictionary. In this setting, a dictionary is obtained by training over training data signals $\boldsymbol{Y} \in \mathbb{C}^{N\times L}$ \cite{ksvd}. This dictionary learning process can be formulated as
\begin{equation}
\argminl_{\boldsymbol{W, D}} {\|\boldsymbol{W}_i\|}_0 ~ s.t. ~ {\|\boldsymbol{Y}_i- \boldsymbol{DW}_i\|}_2^2 < \epsilon ~\forall~ ~i, \label{equa3}
\end{equation}
\noindent where $\epsilon$ represents error tolerance. Since the problem is non-tractable and non-convex, most of the dictionary learning algorithms perform the learning by iteratively alternating between a sparse representation stage and a dictionary update stage. As an example, the K-SVD algorithm \cite{ksvd} is one of the widely used algorithms for the dictionary learning process.

\par The above-mentioned dictionary learning is a computationally demanding process. Therefore, developing efficient alternatives to the classical dictionary learning approach is needed for CR-related applications \cite{choi2017compressed}. In this context, the use of sampled dictionaries is an efficient alternative. One can obtain a sampled dictionary by picking a set of randomly-selected data vectors that serve for the sparse coding without the need for applying an expensive learning process. Thus, this offers a compromise in terms of computational complexity at a tolerable loss in the representational power of the dictionary. In \cite{8688400}, the use of sampled dictionaries is justified by their usage to represent data points in a specific class, which have a general similarity. Similarly, sampled dictionaries are used in this work to represent signals.

\subsection{Residual Components in Pursuit Sparse Coding}
\par A widely used sparse representation algorithm is OMP. This algorithm is based on iteratively obtaining the coefficients in a sparse coefficient vector ($\boldsymbol{w}$). Particularly, each iteration identifies the location and adjusts the value of a nonzero element in $\boldsymbol{w}$. This is achieved by selecting one atom (column) from a dictionary $\boldsymbol{D}$ and adjusting its respective weight.

\par To implement the above-explained atom selection and coefficient update processes, algorithms such as OMP define a so-called residual signal $\boldsymbol{r}$. Conceptually, $\boldsymbol{r}$ represents signal portions that have not yet been represented by the selected dictionary atoms. Hence, sparse coding initializes $\boldsymbol{r}$ with the signal itself, as $\boldsymbol{r} \gets \boldsymbol{x}$. In the first iteration, the sparse representation algorithm loops through all dictionary atoms and selects the one most similar to the current residual $\boldsymbol{r}$. Once this atom is selected, the corresponding weight is calculated. To this end, the next residual is calculated by subtracting the resultant one-atom sparse approximation from the original residual. Then, the residual is considered as a new signal for which another dictionary atom is selected and another coefficient is calculated and the process continues until a certain halting condition is met.

\par The interesting point to consider in the above-explained sparse coding approach is that the energy of the residual components should dramatically decrease as sparse coding progresses. Intuitively, this is because more atoms are selected, and thus more signal portions are excluded from the residual.

\subsection{Machine Learning for Classification} 

\par The successful works of the ML algorithms in many application areas such as computer vision, fingerprint identification, image processing, and speech recognition led these algorithms to become appealing for the area of wireless communication \cite{machinelearning1}. These ML algorithms are categorized under three categories called supervised, unsupervised, and reinforcement learning. Supervised learning-based ML algorithms are widely used for classification problems when the number of present classes is known and the information of the classes that samples belong to in the training stage is available.

\par Amongst many supervised learning-based algorithms, the feed-forward neural network has received growing interest in classification problems since it can recognize classes accurately and quickly \cite{matlab}. This network can be used with a single-layer and multi-layer. Although single-layer algorithms are computationally good, these algorithms can only be used for simple problems. Alternatively, the multi-layer-based algorithms that include the usage of one or more hidden layers are used. Even though these algorithms increase computational complexity, they are able to solve more complex problems. Besides the effect of the extra layers, the number of neurons that are used in hidden layers is also effective on the accuracy and complexity performances. Therefore, it is quite significant to set these hyper-parameters optimally. Moreover, the complexity and accuracy performances can be increased by feature extraction (with the domain knowledge). Along this line, CS is used to extract features in this work with the aim of increasing the performance of the ML.

\subsection{System Model} 
\par
\begin{figure}[h!]
\centering
\resizebox{0.6\columnwidth}{!}{
\includegraphics[width=11cm]{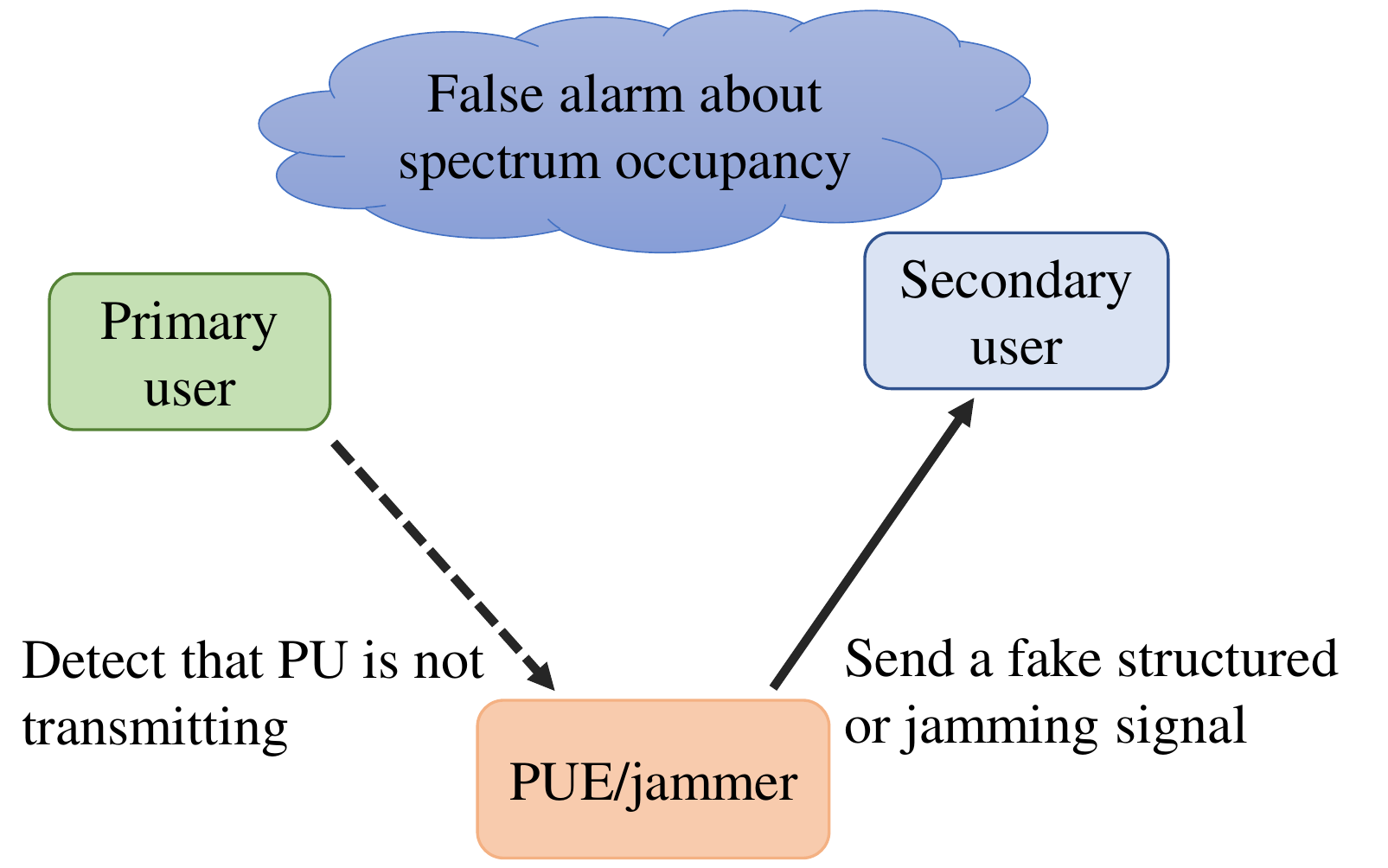}}
\vspace{-8pt}
\linespread{1}
\caption{The basic system model: a PUE and a jammer want to degrade SU's spectrum utilization by sending fake signals.}
\vspace{-4pt}
\label{fig1}
\end{figure}

 The system model used is intended to characterize the existence of legitimate and illegitimate source nodes. Thus, it consists of a PU node, an SU node, and an illegitimate node as presented in Fig.~\ref{fig1}. In this setting, an SU node opportunistically exploits the spectrum in the presence of an illegitimate node that can launch either PUEA or jamming attack. A jammer transmits a random signal, while a PU node and a PUE transmit structured signals that mimic the legitimate PUs.

\par We can represent the transmitted signal as:
$\boldsymbol{x}=\boldsymbol{A}\boldsymbol{s}$, where $\boldsymbol{A}$ is a coefficient matrix with a size of $N \times N$. Each component is denoted by $a_{i,j}$ with $i, j = 1, \ldots N$, and $\boldsymbol{s}=[ s_1(t), \ldots, s_N(t) ]^T$ represents the transmitted data vector. Any coordinate of $\boldsymbol{s}$ is given as $s_i(t)= \sum^\infty_{k=-\infty } d_k u(t-k T_s) e^{j2\pi f_{c,o}t}$, where $T_s$ is the symbol duration, $f_{c,o}$ represents the center frequency, $d$ represents digitally modulated data symbols, $u(t)$ represents the pulse shaping filter, and $o=1,2, \ldots, N$.

\par The signal at the receiver sent by any node can be written as
\begin{equation}
\boldsymbol{y}=\boldsymbol{h}\boldsymbol{x} +\boldsymbol{n},\label{equa5}
\end{equation}
\noindent where $\boldsymbol{h}$ is a multipath Rayleigh fading channel between any transmitter-receiver pair and $\boldsymbol{n}$ is additive white Gaussian noise. Due to the spatial decorrelation concept, the channel between different transmitter-receiver pairs is assumed to be different \cite{8509094}. 

\section{The Proposed Algorithm for PUEA and Jamming Attack Detection}
\label{Section3}
\par The objective of this work is to differentiate between the following hypotheses: 
\begin{equation}
\boldsymbol{y}=
\left\{\begin{array}{ll}
\boldsymbol{n} & \mathcal{H}_0: \text{there is no PU}, \\
\boldsymbol{h}_{PU}\boldsymbol{x_s}+\boldsymbol{n} & \mathcal{H}_1 : \text{a PU is present},\\
\boldsymbol{h}_{i}\boldsymbol{x_s}+\boldsymbol{n} & 
\mathcal{H}_2 : \text{a PUE is present},\\
\boldsymbol{h}_{i}\boldsymbol{x_n}+\boldsymbol{n} & 
\mathcal{H}_3 : \text{a jammer is present},
\label{equa8}
\end{array} \right.
\end{equation}
\noindent where $\boldsymbol{n}$ is additive white Gaussian noise and $\boldsymbol{y}$ is the received signal. Besides, $\boldsymbol{h}_{PU}$ denotes the channel corresponding to the legitimate PU, $\boldsymbol{h}_{i}$ is the channel corresponding to PUE or jammer, $\boldsymbol{x}_{n}$ represents the (unstructured) jamming signal, and $\boldsymbol{x}_{s}$ is a structured signal. In this work, two goals are set. The first is to detect PUEA, i.e., to differentiate between the $\mathcal{H}_0$, $\mathcal{H}_1$, and $\mathcal{H}_2$ hypotheses. The second goal is to detect jamming attacks, i.e., to differentiate between $\mathcal{H}_0$, $\mathcal{H}_1$, and $\mathcal{H}_3$.

\par To meet the above-mentioned goals, a compressed version of the received signal is observed by the CS algorithm and its sparse coding is calculated with respect to a PU channel-depended dictionary $\boldsymbol{D}_{PU}$. As detailed in Section \ref{Section2}, sparse coding iteratively minimizes the energy of a residual ($\|\boldsymbol{r}\|_2$). For each iteration, we calculate the value of $\|\boldsymbol{r}\|_2$. Then, we quantify the rate of its decay using the gradient operator ($|\boldsymbol{G}|$). It is noted that the speed of this decay depends on the harmony between the received signal and the dictionary.

\par The convergence profile of this residual or gradient versus iteration can be used to distinguish between the aforementioned hypotheses. The idea behind this approach is that the unstructured signals (noise and jamming) are not compressible, while structured signals are compressible. Hence, different signals have different $\|\boldsymbol{r}\|_2$ and $|\boldsymbol{G}|$ profiles that help to distinguish between different hypotheses. Following the same logic, different signals have different patterns based on the similarity between the dictionary atoms and signals. In other words, residual energy patterns show how much dictionary atoms can guarantee accurate and sparse representation for signals that can also help in distinguishing between various hypotheses. Intuitively speaking, a signal that is compressible in the given dictionary has a faster decay speed compared to other signals. Thus, if the dictionary is channel-dependent, it will also affect the pattern corresponding to $\|\boldsymbol{r}\|_2$ and $|\boldsymbol{G}|$, which can be used also to differentiate between different hypotheses.

\begin{figure*}[t!]
\centering
\resizebox{0.99\textwidth}{!}{
\begin{tabular}{cccc cccc}
\includegraphics[width=10in,height=6in]{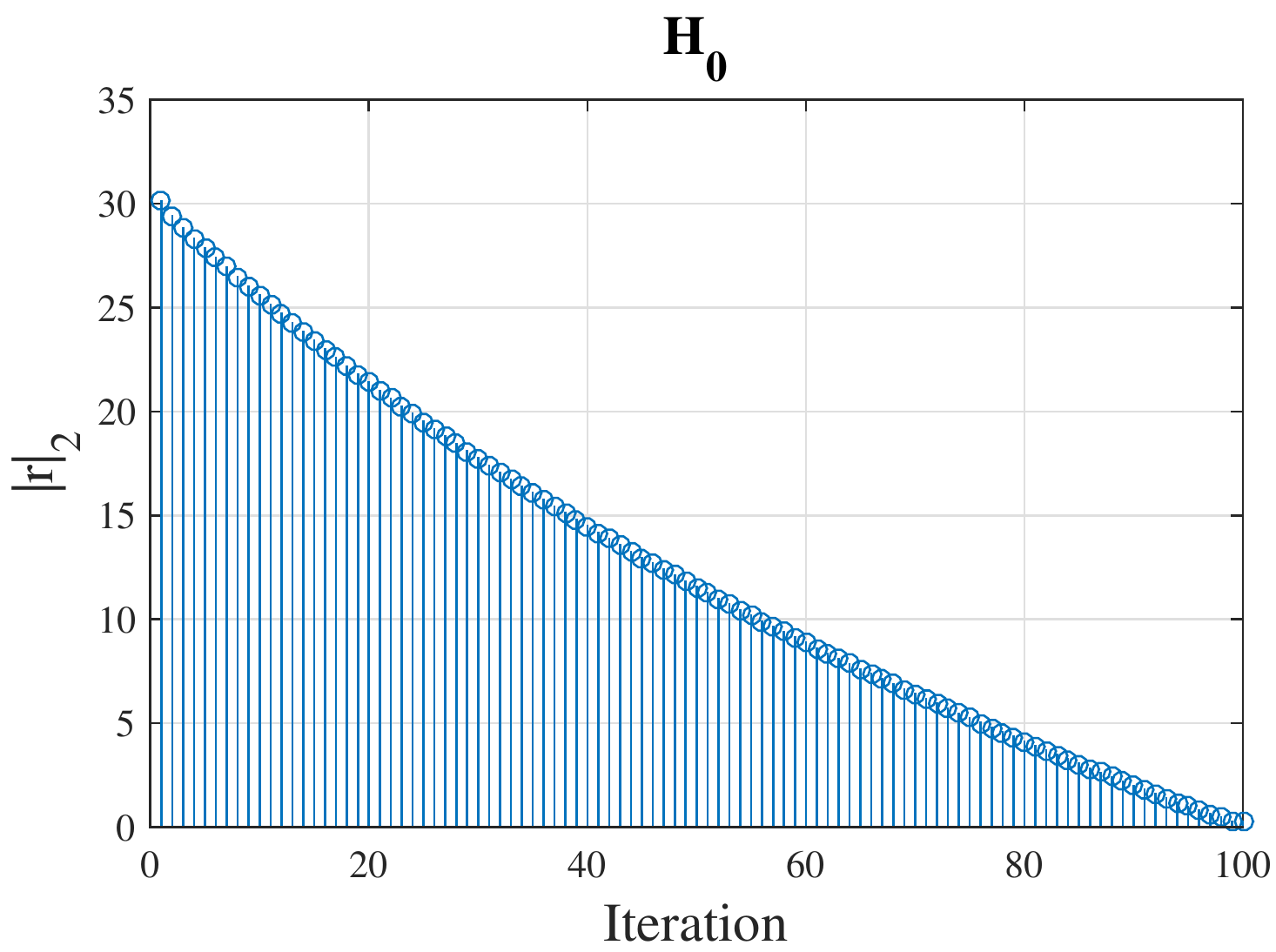}&
\includegraphics[width=10in,height=6in]{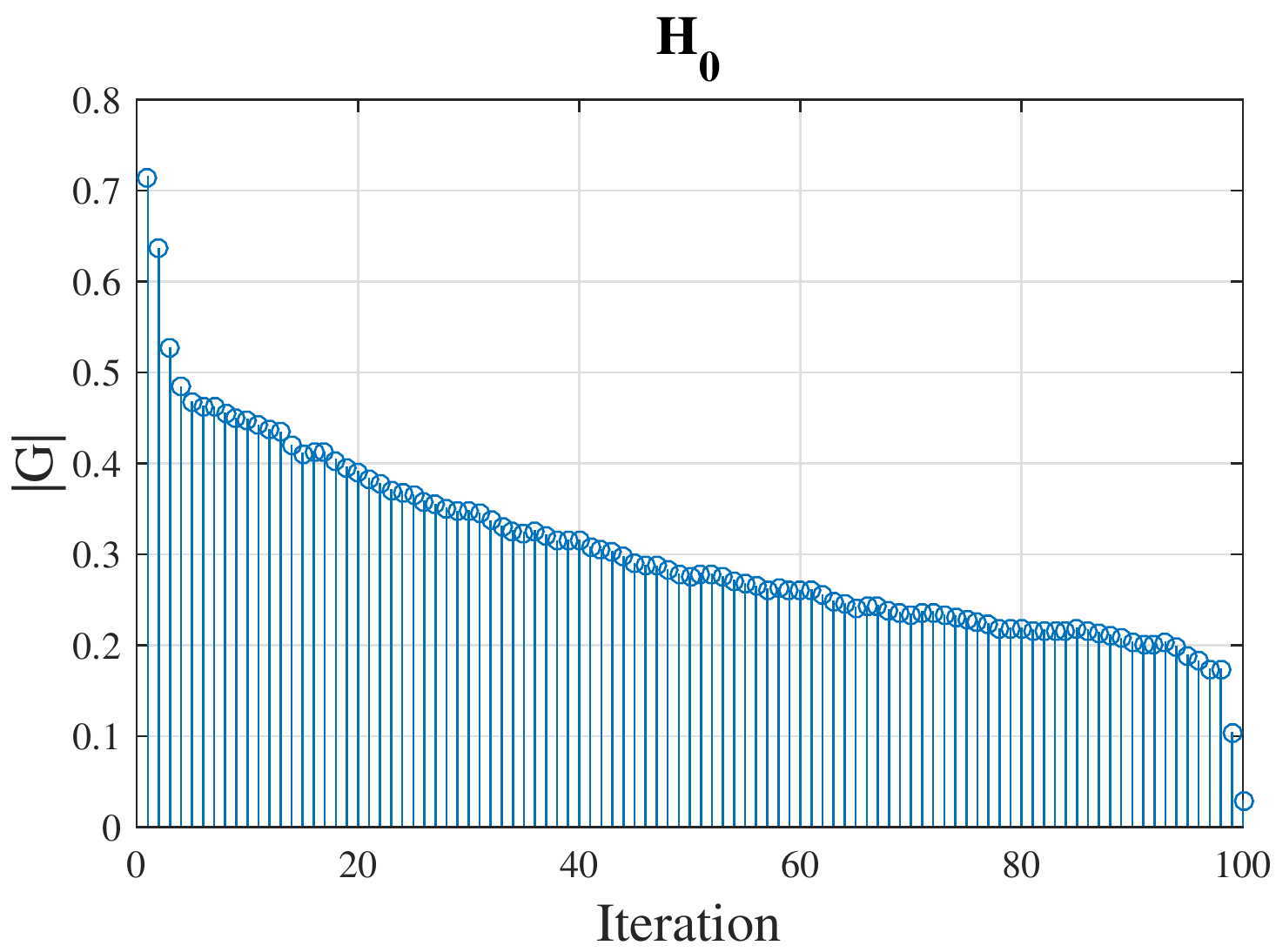}&
\\
\hspace{2.2cm}\Huge(a)&\hspace{1.3cm}\Huge(e)&
\\
\includegraphics[width=10in,height=6in]{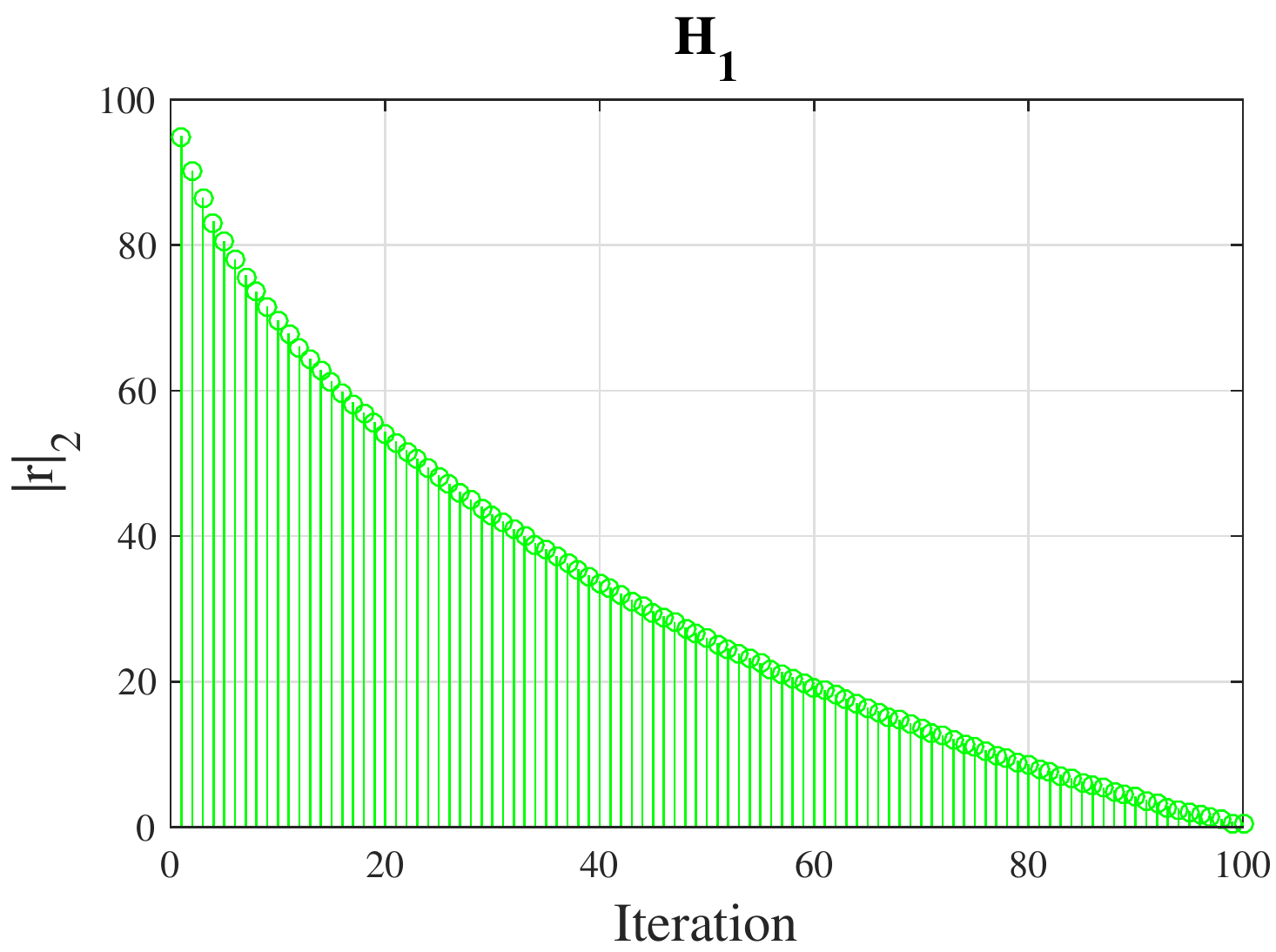}&
\includegraphics[width=10in,height=6in]{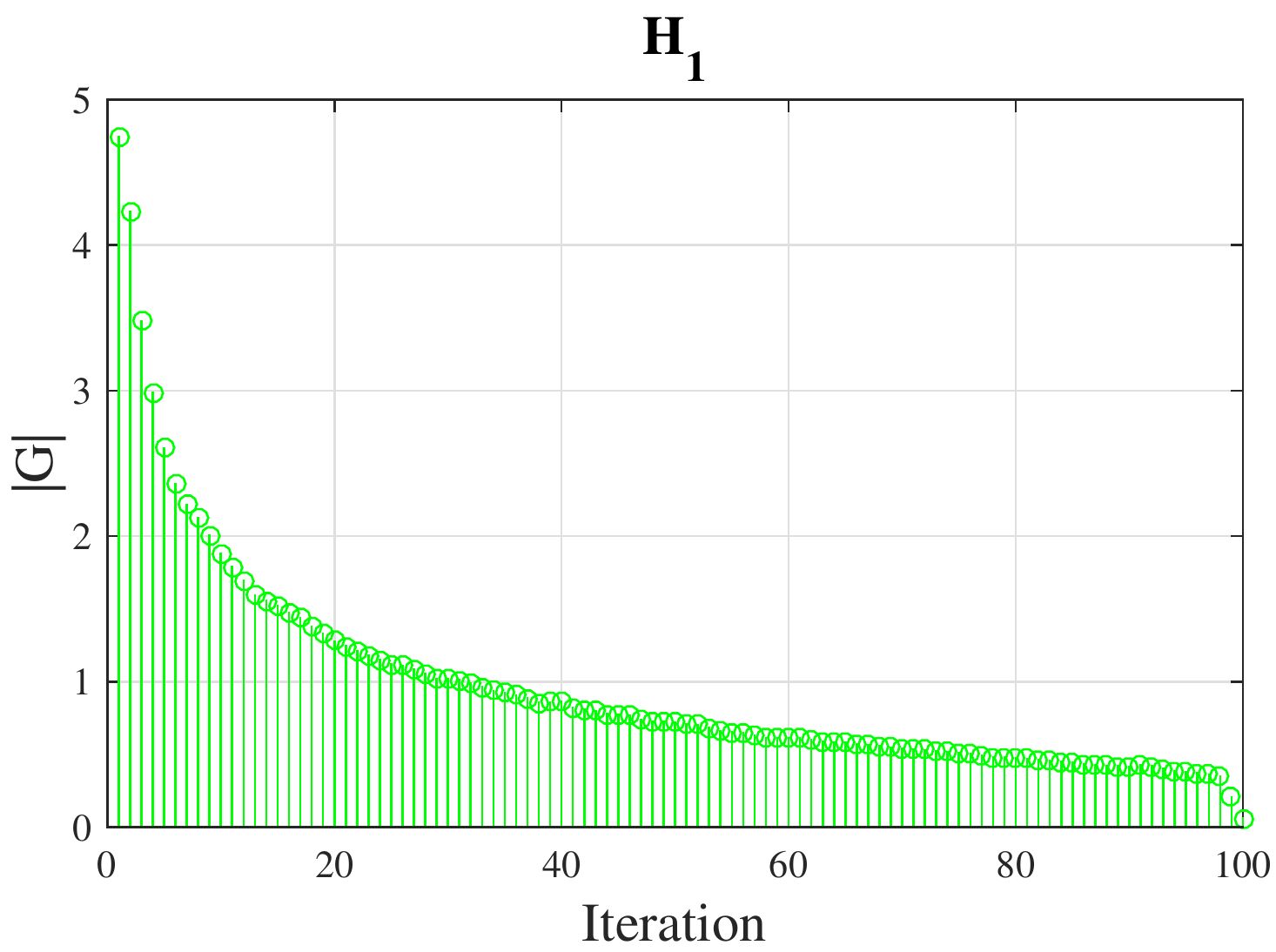}&
\\
\hspace{2.2cm}\Huge(b)&\hspace{1.3cm}\Huge(f)&
\\
\includegraphics[width=10in,height=6in]{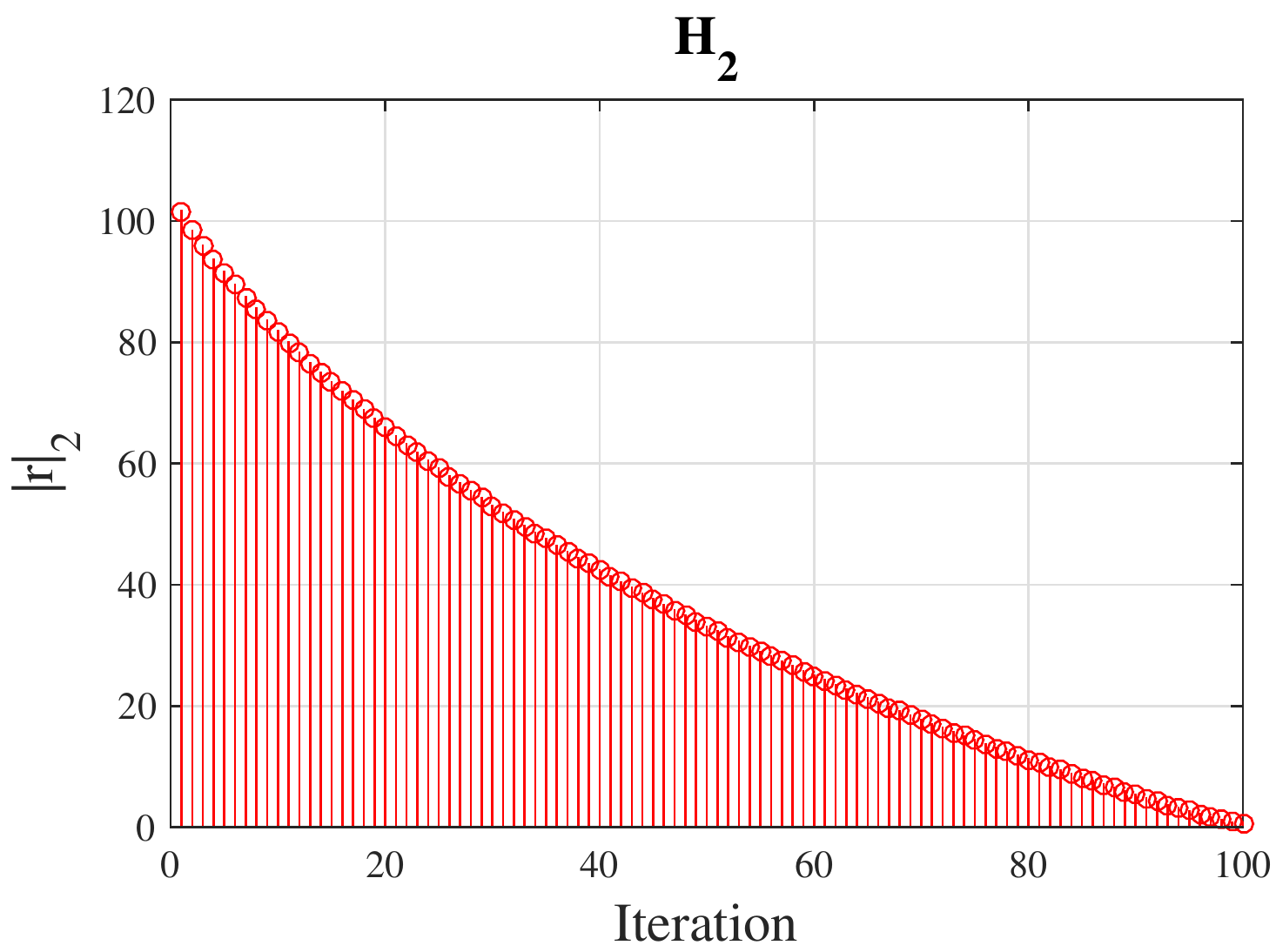}&
\includegraphics[width=10in,height=6in]{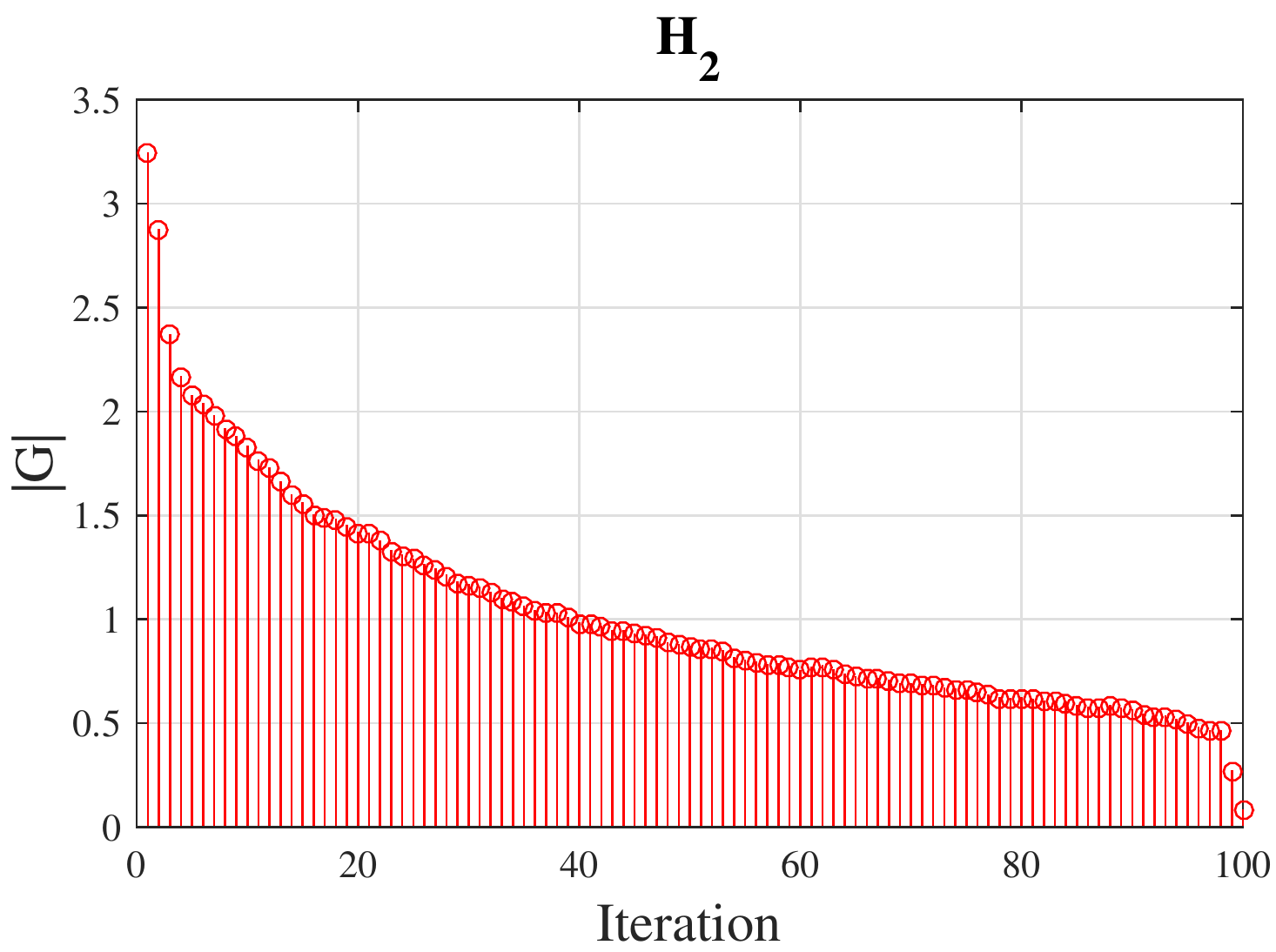}&
\\
\hspace{2.2cm}\Huge(c)&\hspace{1.3cm}\Huge(g)&
\\
\includegraphics[width=10in,height=6in]{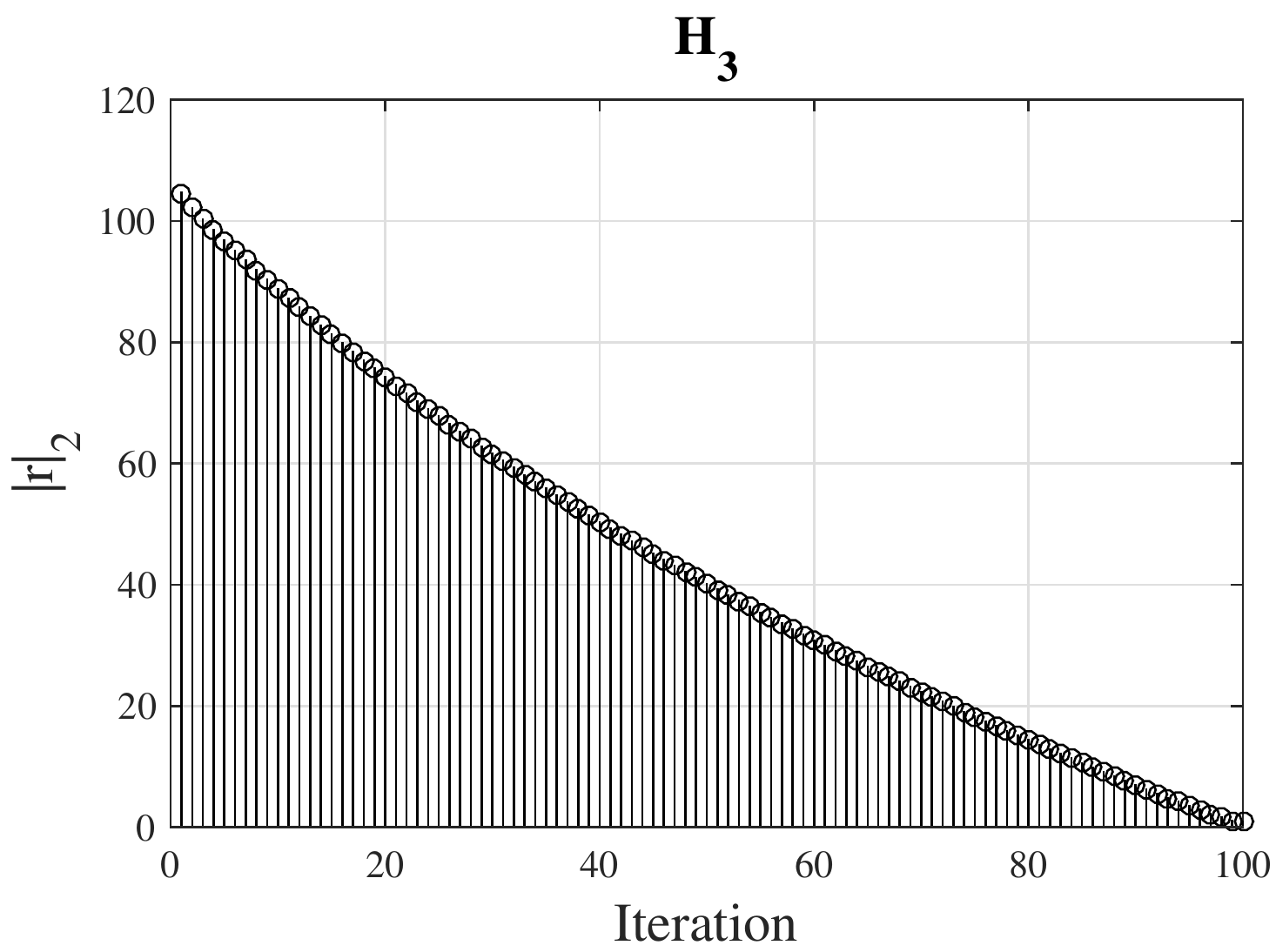}&
\includegraphics[width=10in,height=6in]{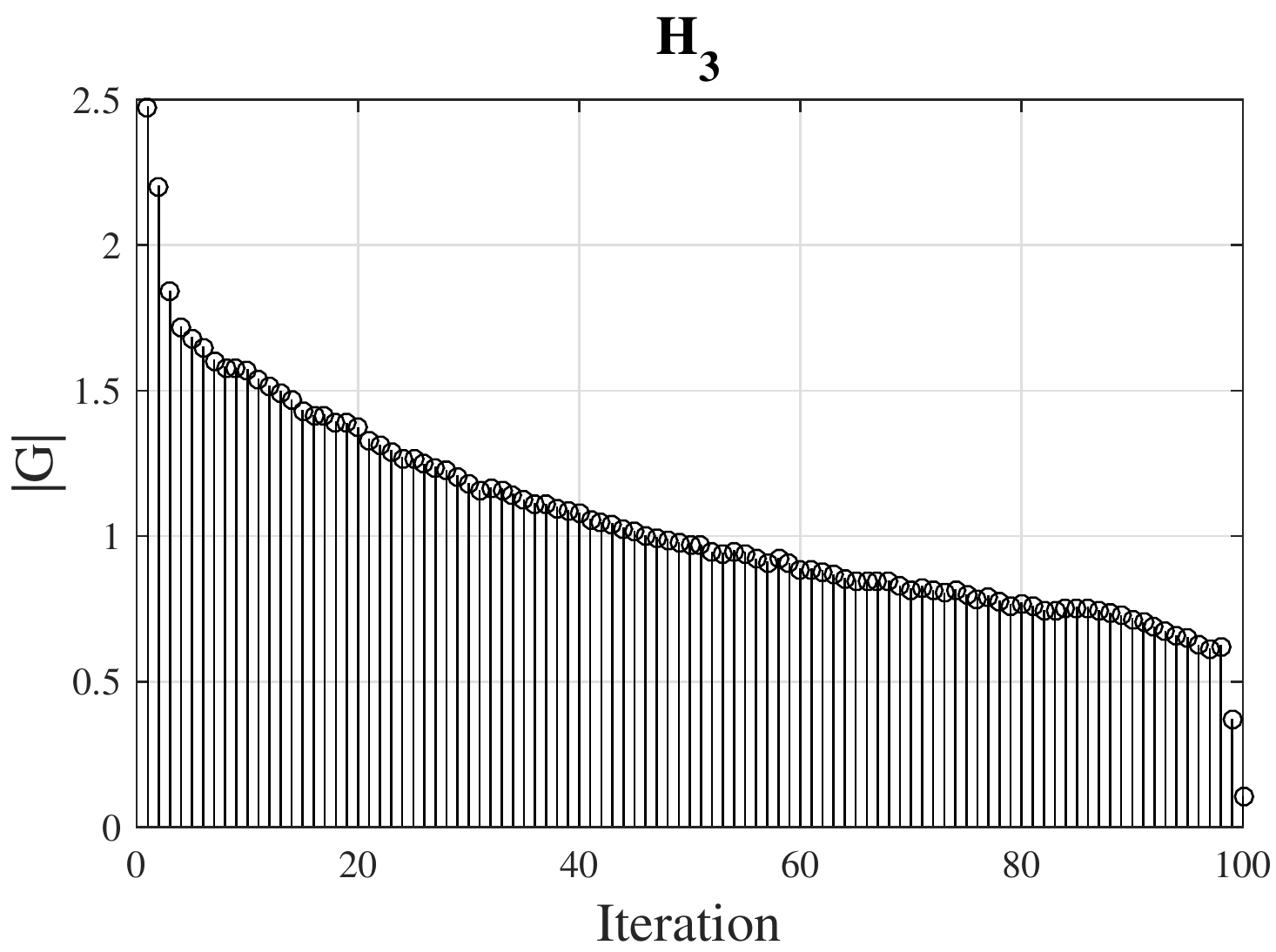}&
\\
\hspace{2.2cm}\Huge(d)&\hspace{1.3cm}\Huge(h)&
\\
\end{tabular}}
\caption{The averages of $\|\boldsymbol{r}\|_2$ versus sparse coding iteration for received signals under hypotheses $\mathcal{H}_0$, $\mathcal{H}_1$, $\mathcal{H}_2$ and $\mathcal{H}_3$ are in (a), (b), (c), and (d), respectively, while the averages of $|\boldsymbol{G}|$ versus sparse coding iteration are presented in (e), (f), (g), and (h), respectively.}
\label{fig3} 
\end{figure*}

\par To this end, we analyze the usefulness of $\|\boldsymbol{r}\|_2$ and $|\boldsymbol{G}|$ in distinguishing between the aforementioned hypotheses in (\ref{equa8}) with the following test. We use a test set of $10^3$ quadruplets of synthetically-generated received signals ($\boldsymbol{y}$) that correspond to the hypotheses $\mathcal{H}_0, \mathcal{H}_1$, $\mathcal{H}_2$, and $\mathcal{H}_3$, respectively. In other words, one signal is mere noise, the other one is the signal received from the legitimate PU, the third one is a PUE signal that mimics the PU signal, and the fourth one is an unstructured jamming signal. These signals are generated as described in Section \ref{Section5}. 

\par For each quadruplet, we calculate a PU-dependent dictionary ($\boldsymbol{D}_{PU}$) based on the known PU channel ($\boldsymbol{h}_{PU}$). In this work, a channel-dependent dictionary is obtained by convolving a set of randomly selected data ($\boldsymbol{X}$) with the channel corresponding to the legitimate PU. Formally stated, $\boldsymbol{D}_{PU}=\boldsymbol{h}_{PU}*\boldsymbol{X}$, where $*$ denotes convolution. Afterwards, we perform an iterative sparse coding operation on a compressed version of each signal in the quadruplet with $\boldsymbol{D}_{PU}$ while calculating $\|\boldsymbol{r}\|_2$. Next, we calculate the gradient of each residual vector as $|\boldsymbol{G}|$. 

\par The average values of $\|\boldsymbol{r}\|_2$ and $|\boldsymbol{G}|$ in the above-explained test are presented in Fig.~\ref{fig3}. In view of this figure, it is seen that one can differentiate between the four hypotheses based on $|\boldsymbol{G}|$ and $\|\boldsymbol{r}\|_2$ using ML approaches. For example, the gradient of $\mathcal{H}_1$ has faster decay as compared to $\mathcal{H}_0$, $\mathcal{H}_2$, $\mathcal{H}_3$ as presented in Fig. 2 (f), Fig. 2 (e), Fig. 2 (g), and Fig. 2 (h), respectively. The reason for exhibiting a faster decay is that the received signal in $\mathcal{H}_1$ (corresponding to PU) is the only one compressible in the given dictionary. 

\par Based on the above discussion, we present the proposed algorithm. It is divided into two main stages. First, is a classifier training stage, where one uses a comprehensive set of training signals. We can either concatenate $\|\boldsymbol{r}\|_2$ and its absolute gradient $|\boldsymbol{G}|$ into a unified feature vector or use them separately as classification features. These features are used to make training data sets $\boldsymbol{f}_0^i$, $\boldsymbol{f}_1^i$, $\boldsymbol{f}_2^i$, and $\boldsymbol{f}_3^i$ according to the hypotheses explained in (\ref{equa8}).

\par
\begin{figure}[!t]
\centering
\begin{tabular}{@{}c@{}}
\includegraphics[width=.5\linewidth]{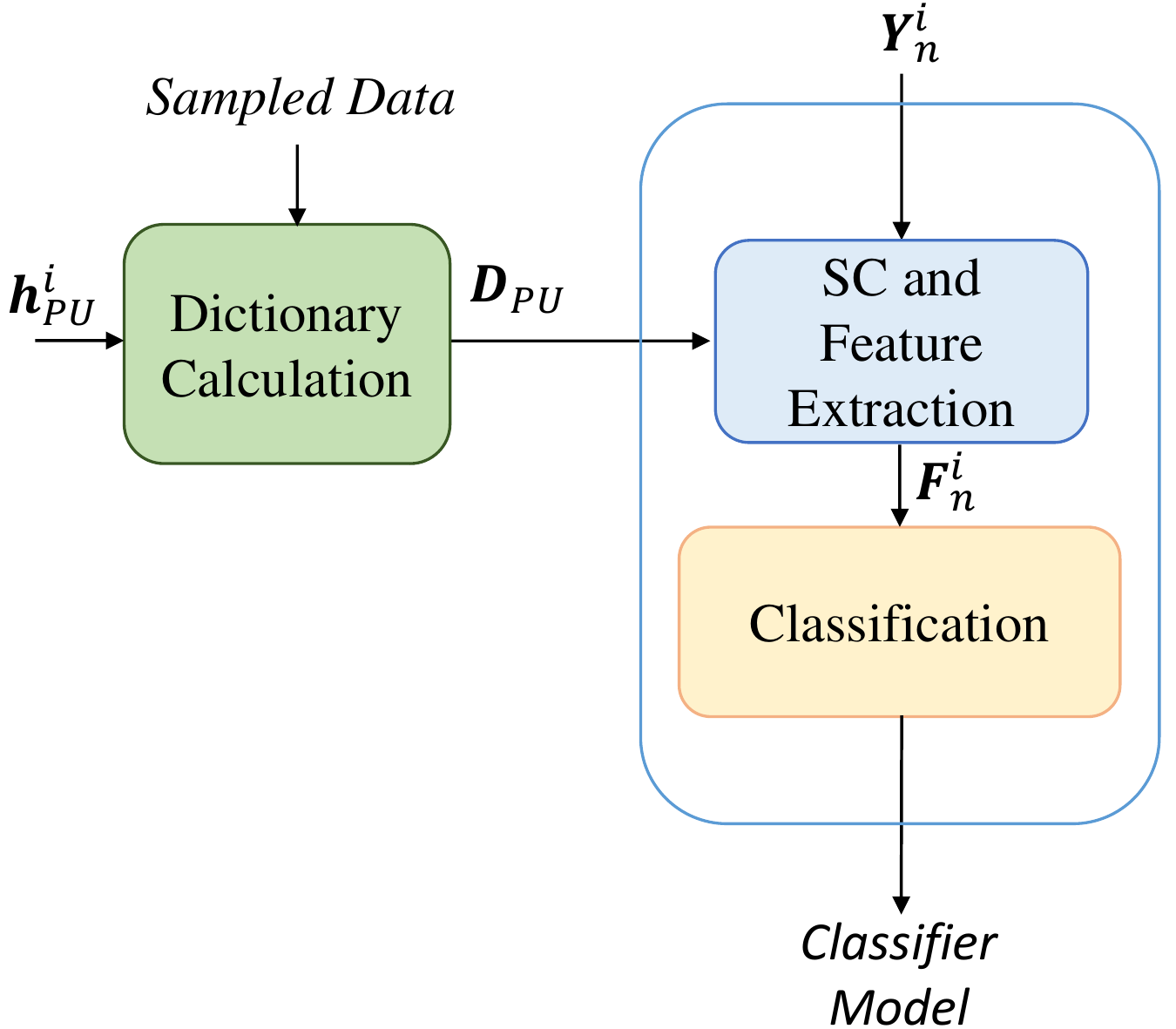} \\
\small \hspace{3.36cm}(a) 
\end{tabular}
\begin{tabular}{c} 
\includegraphics[width=.365\linewidth]{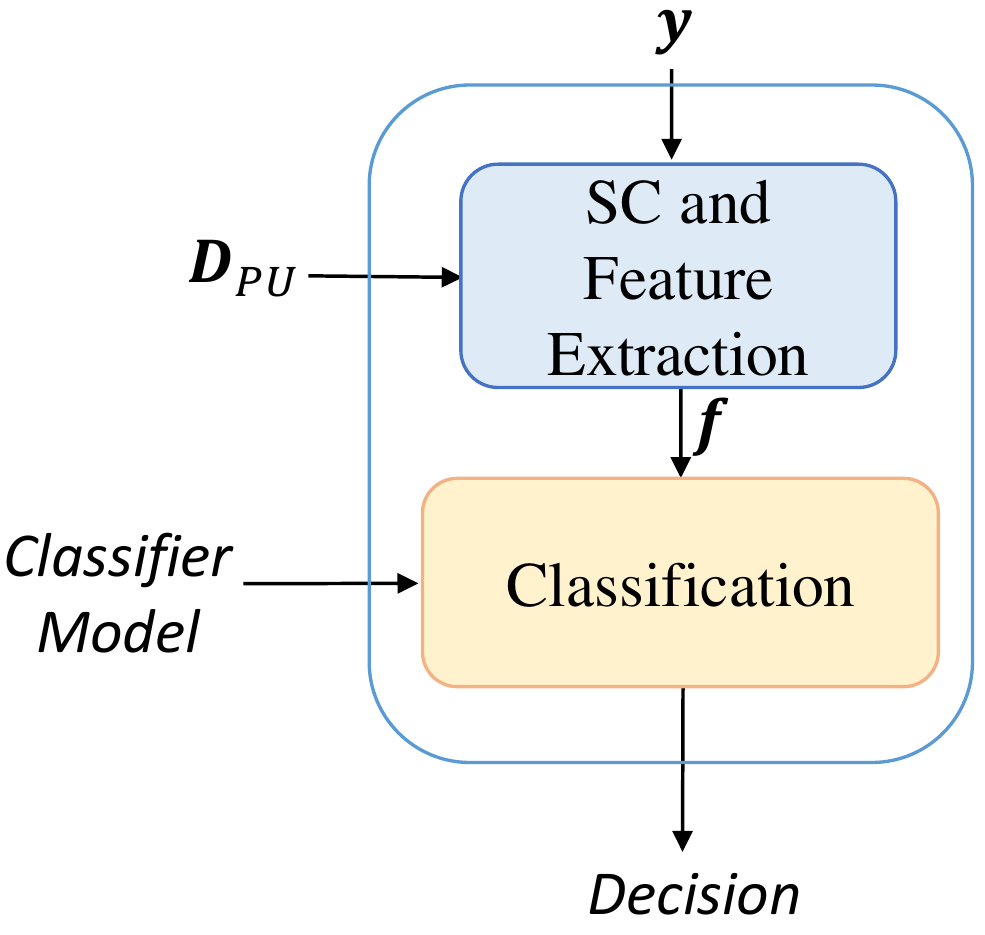} \\
\small \vspace{-1.2cm}\hspace{1.85cm}(b) 
\end{tabular}
\caption{An illustration of the proposed algorithm for (a) training stage, (b) testing stage.}
\label{proposed}
\end{figure}

 For the case of PUEA detection, the training set contains $\boldsymbol{f}_0^i$, $\boldsymbol{f}_1^i$, and $\boldsymbol{f}_2^i$ corresponding to the hypotheses $\mathcal{H}_0, \mathcal{H}_1$, and $\mathcal{H}_2$, respectively. On the other hand, for the case of jammer detection, the training set contains $\boldsymbol{f}_0^i$, $\boldsymbol{f}_1^i$, and $\boldsymbol{f}_3^i$ corresponding to the hypotheses $\mathcal{H}_0, \mathcal{H}_1$ and $\mathcal{H}_3$, respectively. Afterwards, these training vectors, along with their class labels are fed to the ML training stage, where a classifier model is trained accordingly. The workflow of the training set preparation stage is pictorially described in Fig.~\ref{proposed}-(a). In this figure, ${\textbf{Y}}_n^i$ represents the set of compressed received signals $\boldsymbol{y}_0^i$, $\boldsymbol{y}_1^i$, $\boldsymbol{y}_2^i$ for the case of PUEA detection or $\boldsymbol{y}_0^i$, $\boldsymbol{y}_1^i$, $\boldsymbol{y}_3^i$ for the case of jamming attack detection. Similarly, ${\textbf{F}}_n^i$ represents the set of training vectors.

\par After classifier training, the testing stage represents the run-time operation of the proposed algorithm. This process is explained in Fig.~\ref{proposed}-(b). For each incoming test signal, $\boldsymbol{y}$, sparse coding is performed over $\boldsymbol{D}_{PU}$ and feature vector $\boldsymbol{f}$ is obtained. Afterwards, $\boldsymbol{f}$ is fed into the learned classifier. Finally, this classifier will decide on the hypothesis corresponding to the current signal of interest. An analysis of this idea is provided in the Appendix.

\section{Complexity Analysis} 
\label{Section4}

\par In this section, we roughly quantify the computational complexity of the proposed algorithm. This complexity is primarily required by sparse coding and ML.

\par The OMP computational complexity at the $k\mbox{-}$th iteration is $\bigO(MK+KS+KS^2+S^3)$ while the overall complexity is $\bigO(MKS+KS^2+KS^3+S^4)$, where $S$ represents the sparsity level \cite{ompcomp}. Thus, the overall computational complexity of sparse coding with a sparsity level of $M$ is $\bigO(KM^2+KM^2+KM^3+M^4)$. This can be simplified as $\bigO(2KM^2+KM^3+M^4)$. Note that sparse coding is used during both the training and the testing phase in the proposed algorithm.

\par The computational complexity of ML is divided into two main stages which are training and testing. The computational complexity of two-layer neural network per sample is $\bigO(e(lk+ml))$ for training stage, where $e$ denotes the number of epochs, while $k$, $l$, and $m$ represent the number of neurons at the input, hidden, and output layers, respectively. The total complexity of training stage is $\bigO(ep(lk+ml))$ for $p$ number of samples. Moreover, the computational complexity of training per sample is roughly double as compared to the complexity of testing per sample \cite{2014arXiv1412.1710H}. It is worth to note that $k=2M$, since $\|\boldsymbol{r}\|_2$ and $|\boldsymbol{G}|$ are concatenated into a unified feature vector in the simulations.
\section{Results and Discussion}
\label{Section5}
\par This section presents numerical experiments to assess the performance of the proposed algorithm comparing it with the ED approach.

\subsection{Parameter Setting}

\par The simulations are conducted with different modulation settings based on the system model specifications presented in Section \ref{Section2}. The modulation types used include quadrature amplitude modulation (QAM), pulse amplitude modulation (PAM), frequency-shift keying (FSK), and phase-shift keying (PSK). Moreover, the proposed algorithm uses a 100 $\times$ 400 dictionary. For each received signal, a channel realization \cite{2} is generated for the PU and uncorrelated channel realizations are generated for illegitimate node based on channel decorrelation concept \cite{8509094}. The assumed model of $\boldsymbol{h}_{PU}$ is: $\boldsymbol{h}_{PU}=\rho \boldsymbol{h} + (1-\rho)$, where $\rho$ is the correlation factor and $\boldsymbol{h}$ is Rayleigh fading channel \cite{furqan2018adaptive}. The details of the simulation parameters are presented in Table \ref{table11}. 

\par We use a standard two-layer feed-forward network \cite{matlab} for that consists of a hidden layer and an output layer with sigmoid functions. The number of hidden neurons is set to $64$ while the number of output neurons is set to the number of elements in the target vector which is $3$ (corresponding to the number of classes in PUEA or jamming attack detection). For the case of PUEA detection, the vectors $\boldsymbol{f}_0^i$, $\boldsymbol{f}_1^i$, and $\boldsymbol{f}_2^i$ are used for training. For jamming attack detection, $\boldsymbol{f}_0^i$, $\boldsymbol{f}_1^i$, and $\boldsymbol{f}_3^i$ are used as input vectors. Energy decay rate and gradient vectors $\|\boldsymbol{r}\|_2$ and $|\boldsymbol{G}|$ are used as feature vectors. Here, the dimension of both $\|\boldsymbol{r}\|_2$ and $|\boldsymbol{G}|$ is $1\times M$. Therefore, the feature vector dimension $1\times 2M$.

\par It is noted that we take 4000 samples from each class in the training stage for all cases and 1000 samples from each class in the testing stage for each of the SNR values. Also, the neural network is trained over the SNR values ranging between $-5$ dB and $15$ dB with a step size of $5$ dB.

\begin{table}[t!]
\centering
\caption{Synthetic received signal simulation parameters.}
\begin{tabular}{|c|c|}
\hline
Parameter & Value \\
\hline
Channel Model & Rayleigh \\
\hline
No. of taps & 7 \\

\hline
Channel Delay Unit & Sample Period\\
\hline
Signal Length & 100 \\
\hline
Oversampling Rate & 10 \\
\hline
Pulse Shaping & Square-root-raised-cos. \\
\hline
Raised Cos. Symbol Span & 50 \\
\hline
Raised Cos. Roll-off Factor & 0.2 \\
\hline
Correlation Factor & 0.9 \\
\hline

\end{tabular}
\label{table11}
\end{table}

\begin{table*}[b!]
\caption {Confusion matrices for PUEA detection.} 
\label{title4} 
\centering
\resizebox{.8\columnwidth}{!}{
\begin{tabular}{|c|c|c|c|c|c|c|c|c|}
\hline
\rowcolor[HTML]{9B9B9B}
\multicolumn{9}{|c|}{\cellcolor[HTML]{9B9B9B}{\color[HTML]{333333} \textbf{M=30}}} \\ \hline
\rowcolor[HTML]{C0C0C0}
\multicolumn{3}{|c|}{\cellcolor[HTML]{C0C0C0}{\color[HTML]{333333} \textbf{0 dB}}} & \multicolumn{3}{c|}{\cellcolor[HTML]{C0C0C0}{\color[HTML]{333333} \textbf{5 dB}}} & \multicolumn{3}{c|}{\cellcolor[HTML]{C0C0C0}{\color[HTML]{333333} \textbf{10 dB}}} \\ \hline
\rowcolor[HTML]{EFEFEF}
{\color[HTML]{333333} $\mathcal{H}_0$} & {\color[HTML]{333333} $\mathcal{H}_1$} & {\color[HTML]{333333} $\mathcal{H}_2$} & {\color[HTML]{333333} $\mathcal{H}_0$} & {\color[HTML]{333333} $\mathcal{H}_1$} & {\color[HTML]{333333} $\mathcal{H}_2$} & {\color[HTML]{333333} $\mathcal{H}_0$} & {\color[HTML]{333333} $\mathcal{H}_1$} & {\color[HTML]{333333} $\mathcal{H}_2$} \\ \hline
\cellcolor[HTML]{9AFF99}{\color[HTML]{333333} 18.2} & \cellcolor[HTML]{FD6864}{\color[HTML]{333333} 31.5} & \cellcolor[HTML]{FE0000}{\color[HTML]{333333} 50.3} & \cellcolor[HTML]{34FF34}{\color[HTML]{333333} 98.4} & \cellcolor[HTML]{FFCCC9}{\color[HTML]{333333} 1.6} & \cellcolor[HTML]{FFFFFF}{\color[HTML]{333333} 0.0} & \cellcolor[HTML]{34FF34}{\color[HTML]{333333} 100.0} & \cellcolor[HTML]{FFFFFF}{\color[HTML]{333333} 0.0} & \cellcolor[HTML]{FFFFFF}{\color[HTML]{333333} 0.0} \\ \hline
\cellcolor[HTML]{FFCCC9}{\color[HTML]{333333} 2.5} & \cellcolor[HTML]{9AFF99}{\color[HTML]{333333} 44.4} & \cellcolor[HTML]{FE0000}{\color[HTML]{333333} 53.1} & \cellcolor[HTML]{FFCCC9}{\color[HTML]{333333} 2.6} & \cellcolor[HTML]{67FD9A}{\color[HTML]{333333} 61.1} & \cellcolor[HTML]{FD6864}{\color[HTML]{333333} 36.3} & \cellcolor[HTML]{FFCCC9}{\color[HTML]{333333} 7.0} & \cellcolor[HTML]{67FD9A}{\color[HTML]{333333} 63.5} & \cellcolor[HTML]{FD6864}{\color[HTML]{333333} 29.5} \\ \hline
\cellcolor[HTML]{FD6864}{\color[HTML]{333333} 32.6} & \cellcolor[HTML]{FD6864}{\color[HTML]{333333} 26.8} & \cellcolor[HTML]{9AFF99}{\color[HTML]{333333} 40.6} & \cellcolor[HTML]{FFCCC9}{\color[HTML]{333333} 0.6} & \cellcolor[HTML]{FE0000}{\color[HTML]{333333} 50.4} & \cellcolor[HTML]{9AFF99}{\color[HTML]{333333} 49.0} & \cellcolor[HTML]{FFCCC9}{\color[HTML]{333333} 3.4} & \cellcolor[HTML]{FD6864}{\color[HTML]{333333} 43.1} & \cellcolor[HTML]{67FD9A}{\color[HTML]{333333} 53.5} \\ \hline
\rowcolor[HTML]{9B9B9B}
\multicolumn{9}{|c|}{\cellcolor[HTML]{9B9B9B}{\color[HTML]{333333} \textbf{M=50}}} \\ \hline
\rowcolor[HTML]{C0C0C0}
\multicolumn{3}{|c|}{\cellcolor[HTML]{C0C0C0}{\color[HTML]{333333} \textbf{0 dB}}} & \multicolumn{3}{c|}{\cellcolor[HTML]{C0C0C0}{\color[HTML]{333333} \textbf{5 dB}}} & \multicolumn{3}{c|}{\cellcolor[HTML]{C0C0C0}{\color[HTML]{333333} \textbf{10 dB}}} \\ \hline
\rowcolor[HTML]{EFEFEF}
{\color[HTML]{333333} $\mathcal{H}_0$} & {\color[HTML]{333333} $\mathcal{H}_1$} & {\color[HTML]{333333} $\mathcal{H}_2$} & {\color[HTML]{333333} $\mathcal{H}_0$} & {\color[HTML]{333333} $\mathcal{H}_1$} & {\color[HTML]{333333} $\mathcal{H}_2$} & {\color[HTML]{333333} $\mathcal{H}_0$} & {\color[HTML]{333333} $\mathcal{H}_1$} & {\color[HTML]{333333} $\mathcal{H}_2$} \\ \hline
\cellcolor[HTML]{9AFF99}{\color[HTML]{333333} 36.0} & \cellcolor[HTML]{FD6864}{\color[HTML]{333333} 29.4} & \cellcolor[HTML]{FD6864}{\color[HTML]{333333} 34.6} & \cellcolor[HTML]{34FF34}{\color[HTML]{333333} 99.4} & \cellcolor[HTML]{FFCCC9}{\color[HTML]{333333} 0.6} & \cellcolor[HTML]{FFFFFF}{\color[HTML]{333333} 0.0} & \cellcolor[HTML]{34FF34}{\color[HTML]{333333} 100.0} & \cellcolor[HTML]{FFFFFF}{\color[HTML]{333333} 0.0} & \cellcolor[HTML]{FFFFFF}{\color[HTML]{333333} 0.0} \\ \hline
\cellcolor[HTML]{FFCCC9}{\color[HTML]{333333} 1.4} & \cellcolor[HTML]{67FD9A}{\color[HTML]{333333} 37.4} & \cellcolor[HTML]{FE0000}61.2 & \cellcolor[HTML]{FFCCC9}{\color[HTML]{333333} 2.9} & \cellcolor[HTML]{34FF34}{\color[HTML]{333333} 71.2} & \cellcolor[HTML]{FD6864}{\color[HTML]{333333} 25.9} & \cellcolor[HTML]{FFCCC9}{\color[HTML]{333333} 7.2} & \cellcolor[HTML]{34FF34}{\color[HTML]{333333} 71.6} & \cellcolor[HTML]{FD6864}{\color[HTML]{333333} 21.2} \\ \hline
\cellcolor[HTML]{FD6864}{\color[HTML]{333333} 35.4} & \cellcolor[HTML]{FD6864}{\color[HTML]{333333} 16.3} & \cellcolor[HTML]{9AFF99}{\color[HTML]{333333} 48.3} & \cellcolor[HTML]{FFCCC9}{\color[HTML]{333333} 0.3} & \cellcolor[HTML]{FE0000}{\color[HTML]{333333} 52.8} & \cellcolor[HTML]{9AFF99}{\color[HTML]{333333} 46.9} & \cellcolor[HTML]{FFCCC9}{\color[HTML]{333333} 5.9} & \cellcolor[HTML]{FD6864}{\color[HTML]{333333} 45.3} & \cellcolor[HTML]{9AFF99}{\color[HTML]{333333} 48.8} \\ \hline
\rowcolor[HTML]{9B9B9B}
\multicolumn{9}{|c|}{\cellcolor[HTML]{9B9B9B}{\color[HTML]{333333} \textbf{M=70}}} \\ \hline
\rowcolor[HTML]{C0C0C0}
\multicolumn{3}{|c|}{\cellcolor[HTML]{C0C0C0}{\color[HTML]{333333} \textbf{0 dB}}} & \multicolumn{3}{c|}{\cellcolor[HTML]{C0C0C0}{\color[HTML]{333333} \textbf{5 dB}}} & \multicolumn{3}{c|}{\cellcolor[HTML]{C0C0C0}{\color[HTML]{333333} \textbf{10 dB}}} \\ \hline
\rowcolor[HTML]{EFEFEF}
{\color[HTML]{333333} $\mathcal{H}_0$} & {\color[HTML]{333333} $\mathcal{H}_1$} & {\color[HTML]{333333} $\mathcal{H}_2$} & {\color[HTML]{333333} $\mathcal{H}_0$} & {\color[HTML]{333333} $\mathcal{H}_1$} & {\color[HTML]{333333} $\mathcal{H}_2$} & {\color[HTML]{333333} $\mathcal{H}_0$} & {\color[HTML]{333333} $\mathcal{H}_1$} & {\color[HTML]{333333} $\mathcal{H}_2$} \\ \hline
\cellcolor[HTML]{67FD9A}{\color[HTML]{333333} 57.6} & \cellcolor[HTML]{FD6864}{\color[HTML]{333333} 19.1} & \cellcolor[HTML]{FD6864}{\color[HTML]{333333} 23.3} & \cellcolor[HTML]{34FF34}{\color[HTML]{333333} 99.9} & \cellcolor[HTML]{FFCCC9}{\color[HTML]{333333} 0.1} & {\color[HTML]{333333} 0.0} & \cellcolor[HTML]{34FF34}{\color[HTML]{333333} 100.0} & \cellcolor[HTML]{FFFFFF}{\color[HTML]{333333} 0.0} & \cellcolor[HTML]{FFFFFF}{\color[HTML]{333333} 0.0} \\ \hline
\cellcolor[HTML]{FFCCC9}{\color[HTML]{333333} 1.5} & \cellcolor[HTML]{34FF34}{\color[HTML]{333333} 70.9} & \cellcolor[HTML]{FD6864}{\color[HTML]{333333} 27.6} & \cellcolor[HTML]{FFCCC9}{\color[HTML]{333333} 4.3} & \cellcolor[HTML]{67FD9A}{\color[HTML]{333333} 57.0} & \cellcolor[HTML]{FD6864}{\color[HTML]{333333} 38.7} & \cellcolor[HTML]{FFCCC9}{\color[HTML]{333333} 8.1} & \cellcolor[HTML]{67FD9A}{\color[HTML]{333333} 69.2} & \cellcolor[HTML]{FD6864}{\color[HTML]{333333} 22.7} \\ \hline
\rowcolor[HTML]{FD6864}
{\color[HTML]{333333} 33.8} & {\color[HTML]{333333} 15.9} & \cellcolor[HTML]{67FD9A}{\color[HTML]{333333} 50.3} & \cellcolor[HTML]{FFCCC9}{\color[HTML]{333333} 0.4} & {\color[HTML]{333333} 43.8} & \cellcolor[HTML]{67FD9A}{\color[HTML]{333333} 55.8} & {\color[HTML]{333333} 10.7} & {\color[HTML]{333333} 34.9} & \cellcolor[HTML]{67FD9A}{\color[HTML]{333333} 54.4} \\ \hline
\rowcolor[HTML]{C0C0C0}
\multicolumn{9}{|c|}{\cellcolor[HTML]{C0C0C0}{\color[HTML]{333333} \textbf{M=100}}} \\ \hline
\rowcolor[HTML]{C0C0C0}
\multicolumn{3}{|c|}{\cellcolor[HTML]{C0C0C0}\textbf{0 dB}} & \multicolumn{3}{c|}{\cellcolor[HTML]{C0C0C0}\textbf{5 dB}} & \multicolumn{3}{c|}{\cellcolor[HTML]{C0C0C0}\textbf{10 dB}} \\ \hline
\rowcolor[HTML]{EFEFEF}
{\color[HTML]{333333} $\mathcal{H}_0$} & {\color[HTML]{333333} $\mathcal{H}_1$} & {\color[HTML]{333333} $\mathcal{H}_2$} & {\color[HTML]{333333} $\mathcal{H}_0$} & {\color[HTML]{333333} $\mathcal{H}_1$} & {\color[HTML]{333333} $\mathcal{H}_2$} & {\color[HTML]{333333} $\mathcal{H}_0$} & {\color[HTML]{333333} $\mathcal{H}_1$} & {\color[HTML]{333333} $\mathcal{H}_2$} \\ \hline
\cellcolor[HTML]{34FF34}{\color[HTML]{333333} 88.9} & \cellcolor[HTML]{FFCCC9}{\color[HTML]{333333} 6.4} & \cellcolor[HTML]{FFCCC9}{\color[HTML]{333333} 4.7} & \cellcolor[HTML]{34FF34}{\color[HTML]{333333} 100.0} & {\color[HTML]{333333} 0.0} & {\color[HTML]{333333} 0.0} & \cellcolor[HTML]{34FF34}{\color[HTML]{333333} 100.0} & {\color[HTML]{333333} 0.0} & {\color[HTML]{333333} 0.0} \\ \hline
\cellcolor[HTML]{FFCCC9}{\color[HTML]{333333} 1.4} & \cellcolor[HTML]{34FF34}{\color[HTML]{333333} 81.2} & \cellcolor[HTML]{FD6864}{\color[HTML]{333333} 17.4} & \cellcolor[HTML]{FFCCC9}{\color[HTML]{333333} 1.0} & \cellcolor[HTML]{67FD9A}{\color[HTML]{333333} 53.0} & \cellcolor[HTML]{FD6864}{\color[HTML]{333333} 46.0} & \cellcolor[HTML]{FFCCC9}{\color[HTML]{333333} 4.0} & \cellcolor[HTML]{34FF34}{\color[HTML]{333333} 79.4} & \cellcolor[HTML]{FD6864}{\color[HTML]{333333} 16.6} \\ \hline
\cellcolor[HTML]{FD6864}{\color[HTML]{333333} 32.2} & \cellcolor[HTML]{FFCCC9}{\color[HTML]{333333} 6.7} & \cellcolor[HTML]{67FD9A}{\color[HTML]{333333} 61.1} & \cellcolor[HTML]{FFCCC9}{\color[HTML]{333333} 0.3} & \cellcolor[HTML]{FD6864}{\color[HTML]{333333} 17.0} & \cellcolor[HTML]{34FF34}{\color[HTML]{333333} 82.7} & \cellcolor[HTML]{FFCCC9}{\color[HTML]{333333} 3.2} & \cellcolor[HTML]{FD6864}{\color[HTML]{333333} 35.0} & \cellcolor[HTML]{67FD9A}{\color[HTML]{333333} 61.8} \\ \hline
\end{tabular}}
\end{table*}

\subsection{Performance Analysis}

\par This section presents the performance analysis of the proposed algorithm in terms of confusion matrices, receiver operating characteristics (ROC) curves and area under ROC (AUROC) curves. For the jamming detection scenario, it is assumed that the illegitimate node broadcasts non-structured signals. On the other hand, it is assumed that PUE signal's parameters are identical to that of PU signal. 

\par To examine the performance of the classification, confusion matrices are often used. They present the number of both correctly and incorrectly classified observations. Thus, diagonal elements present the number of those observations correctly classified while off-diagonal elements indicate the number of incorrectly classified observations. 

\par Table \ref{title4} presents the confusion matrices for the case of PUEA detection for different $M$ and SNR values, where $M$ is the number of samples in the compressed received signal. It is observed from Table \ref{title4} that the overall performance of the proposed algorithm is satisfactory for PUEA detection, especially at high SNR. Besides, the performance also improves with the increase in the values of $M$. Table \ref{title5} presents the confusion matrices for the case of jamming detection for different $M$ and SNR values. It is seen from the table that the classification accuracy based on the proposed algorithm improves with the increase in $M$ and SNR similar to PUEA case. 

\begin{table*}[t!]
\caption {Confusion matrices for jamming attack detection.} \label{title5} \centering\resizebox{.8\columnwidth}{!}{
\begin{tabular}{|c|c|c|c|c|c|c|c|c|}
\hline
\rowcolor[HTML]{9B9B9B} 
\multicolumn{9}{|c|}{\cellcolor[HTML]{9B9B9B}{\color[HTML]{333333} \textbf{M=30}}} \\ \hline
\rowcolor[HTML]{C0C0C0} 
\multicolumn{3}{|c|}{\cellcolor[HTML]{C0C0C0}{\color[HTML]{333333} \textbf{0 dB}}} & \multicolumn{3}{c|}{\cellcolor[HTML]{C0C0C0}{\color[HTML]{333333} \textbf{5 dB}}} & \multicolumn{3}{c|}{\cellcolor[HTML]{C0C0C0}{\color[HTML]{333333} \textbf{10 dB}}} \\ \hline
\rowcolor[HTML]{EFEFEF} 
{\color[HTML]{333333} $\mathcal{H}_0$} & {\color[HTML]{333333} $\mathcal{H}_1$} & {\color[HTML]{333333} $\mathcal{H}_2$} & {\color[HTML]{333333} $\mathcal{H}_0$} & {\color[HTML]{333333} $\mathcal{H}_1$} & {\color[HTML]{333333} $\mathcal{H}_2$} & {\color[HTML]{333333} $\mathcal{H}_0$} & {\color[HTML]{333333} $\mathcal{H}_1$} & {\color[HTML]{333333} $\mathcal{H}_2$} \\ \hline
\cellcolor[HTML]{9AFF99}{\color[HTML]{333333} 12.5} & \cellcolor[HTML]{FD6864}{\color[HTML]{333333} 25} & \cellcolor[HTML]{FE0000}{\color[HTML]{333333} 62.5} & \cellcolor[HTML]{34FF34}{\color[HTML]{333333} 98.4} & \cellcolor[HTML]{FFCCC9}{\color[HTML]{333333} 1.6} & \cellcolor[HTML]{FFFFFF}{\color[HTML]{333333} 0} & \cellcolor[HTML]{34FF34}{\color[HTML]{333333} 100} & \cellcolor[HTML]{FFFFFF}{\color[HTML]{333333} 0} & \cellcolor[HTML]{FFFFFF}{\color[HTML]{333333} 0} \\ \hline
\cellcolor[HTML]{FFCCC9}{\color[HTML]{333333} 0.9} & \cellcolor[HTML]{9AFF99}{\color[HTML]{333333} 44.6} & \cellcolor[HTML]{FE0000}{\color[HTML]{333333} 54.5} & \cellcolor[HTML]{FFCCC9}{\color[HTML]{333333} 2.3} & \cellcolor[HTML]{67FD9A}{\color[HTML]{333333} 54.8} & \cellcolor[HTML]{FD6864}{\color[HTML]{333333} 42.9} & \cellcolor[HTML]{FFCCC9}{\color[HTML]{333333} 3.5} & \cellcolor[HTML]{67FD9A}{\color[HTML]{333333} 58} & \cellcolor[HTML]{FD6864}{\color[HTML]{333333} 38.5} \\ \hline
\cellcolor[HTML]{FD6864}{\color[HTML]{333333} 27.3} & \cellcolor[HTML]{FD6864}{\color[HTML]{333333} 25.6} & \cellcolor[HTML]{9AFF99}{\color[HTML]{333333} 47.1} & \cellcolor[HTML]{FFFFFF}{\color[HTML]{333333} 0} & \cellcolor[HTML]{FD6864}{\color[HTML]{333333} 27.9} & \cellcolor[HTML]{34FF34}{\color[HTML]{333333} 72.1} & \cellcolor[HTML]{FFCCC9}{\color[HTML]{333333} 1.3} & \cellcolor[HTML]{FD6864}{\color[HTML]{333333} 26.2} & \cellcolor[HTML]{34FF34}{\color[HTML]{333333} 72.5} \\ \hline
\rowcolor[HTML]{9B9B9B} 
\multicolumn{9}{|c|}{\cellcolor[HTML]{9B9B9B}{\color[HTML]{333333} \textbf{M=50}}} \\ \hline
\rowcolor[HTML]{C0C0C0} 
\multicolumn{3}{|c|}{\cellcolor[HTML]{C0C0C0}{\color[HTML]{333333} \textbf{0 dB}}} & \multicolumn{3}{c|}{\cellcolor[HTML]{C0C0C0}{\color[HTML]{333333} \textbf{5 dB}}} & \multicolumn{3}{c|}{\cellcolor[HTML]{C0C0C0}{\color[HTML]{333333} \textbf{10 dB}}} \\ \hline
\rowcolor[HTML]{EFEFEF} 
{\color[HTML]{333333} $\mathcal{H}_0$} & {\color[HTML]{333333} $\mathcal{H}_1$} & {\color[HTML]{333333} $\mathcal{H}_2$} & {\color[HTML]{333333} $\mathcal{H}_0$} & {\color[HTML]{333333} $\mathcal{H}_1$} & {\color[HTML]{333333} $\mathcal{H}_2$} & {\color[HTML]{333333} $\mathcal{H}_0$} & {\color[HTML]{333333} $\mathcal{H}_1$} & {\color[HTML]{333333} $\mathcal{H}_2$} \\ \hline
\cellcolor[HTML]{9AFF99}{\color[HTML]{333333} 24.5} & \cellcolor[HTML]{FD6864}{\color[HTML]{333333} 20.9} & \cellcolor[HTML]{FE0000}{\color[HTML]{333333} 54.6} & \cellcolor[HTML]{34FF34}{\color[HTML]{333333} 99.5} & \cellcolor[HTML]{FFCCC9}{\color[HTML]{333333} 0.5} & \cellcolor[HTML]{FFFFFF}{\color[HTML]{333333} 0} & \cellcolor[HTML]{34FF34}{\color[HTML]{333333} 100} & \cellcolor[HTML]{FFFFFF}{\color[HTML]{333333} 0} & \cellcolor[HTML]{FFFFFF}{\color[HTML]{333333} 0} \\ \hline
\cellcolor[HTML]{FFCCC9}{\color[HTML]{333333} 0.9} & \cellcolor[HTML]{67FD9A}{\color[HTML]{333333} 51.9} & \cellcolor[HTML]{FD6864}{\color[HTML]{333333} 47.2} & \cellcolor[HTML]{FFCCC9}{\color[HTML]{333333} 2.4} & \cellcolor[HTML]{67FD9A}{\color[HTML]{333333} 65.3} & \cellcolor[HTML]{FD6864}{\color[HTML]{333333} 32.3} & \cellcolor[HTML]{FFCCC9}{\color[HTML]{333333} 4.6} & \cellcolor[HTML]{67FD9A}{\color[HTML]{333333} 69.7} & \cellcolor[HTML]{FD6864}{\color[HTML]{333333} 25.7} \\ \hline
\cellcolor[HTML]{FD6864}{\color[HTML]{333333} 42.7} & \cellcolor[HTML]{FFCCC9}{\color[HTML]{333333} 14.5} & \cellcolor[HTML]{9AFF99}{\color[HTML]{333333} 42.8} & \cellcolor[HTML]{FFFFFF}{\color[HTML]{333333} 0} & \cellcolor[HTML]{FD6864}{\color[HTML]{333333} 23.5} & \cellcolor[HTML]{34FF34}{\color[HTML]{333333} 76.5} & \cellcolor[HTML]{FFCCC9}{\color[HTML]{333333} 2.6} & \cellcolor[HTML]{FD6864}{\color[HTML]{333333} 20} & \cellcolor[HTML]{34FF34}{\color[HTML]{333333} 77.4} \\ \hline
\rowcolor[HTML]{9B9B9B} 
\multicolumn{9}{|c|}{\cellcolor[HTML]{9B9B9B}{\color[HTML]{333333} \textbf{M=70}}} \\ \hline
\rowcolor[HTML]{C0C0C0} 
\multicolumn{3}{|c|}{\cellcolor[HTML]{C0C0C0}{\color[HTML]{333333} \textbf{0 dB}}} & \multicolumn{3}{c|}{\cellcolor[HTML]{C0C0C0}{\color[HTML]{333333} \textbf{5 dB}}} & \multicolumn{3}{c|}{\cellcolor[HTML]{C0C0C0}{\color[HTML]{333333} \textbf{10 dB}}} \\ \hline
\rowcolor[HTML]{EFEFEF} 
{\color[HTML]{333333} $\mathcal{H}_0$} & {\color[HTML]{333333} $\mathcal{H}_1$} & {\color[HTML]{333333} $\mathcal{H}_2$} & {\color[HTML]{333333} $\mathcal{H}_0$} & {\color[HTML]{333333} $\mathcal{H}_1$} & {\color[HTML]{333333} $\mathcal{H}_2$} & {\color[HTML]{333333} $\mathcal{H}_0$} & {\color[HTML]{333333} $\mathcal{H}_1$} & {\color[HTML]{333333} $\mathcal{H}_2$} \\ \hline
\cellcolor[HTML]{9AFF99}{\color[HTML]{333333} 29.8} & \cellcolor[HTML]{FD6864}{\color[HTML]{333333} 15} & \cellcolor[HTML]{FE0000}{\color[HTML]{333333} 55.2} & \cellcolor[HTML]{34FF34}{\color[HTML]{333333} 99.5} & \cellcolor[HTML]{FFCCC9}{\color[HTML]{333333} 0.5} & {\color[HTML]{333333} 0} & \cellcolor[HTML]{34FF34}{\color[HTML]{333333} 100} & \cellcolor[HTML]{FFFFFF}{\color[HTML]{333333} 0} & \cellcolor[HTML]{FFCCC9}{\color[HTML]{333333} 0} \\ \hline
\cellcolor[HTML]{FFCCC9}{\color[HTML]{333333} 0.2} & \cellcolor[HTML]{67FD9A}{\color[HTML]{333333} 63.3} & \cellcolor[HTML]{FD6864}{\color[HTML]{333333} 36.5} & \cellcolor[HTML]{FFCCC9}{\color[HTML]{333333} 2.1} & \cellcolor[HTML]{67FD9A}{\color[HTML]{333333} 65.6} & \cellcolor[HTML]{FD6864}{\color[HTML]{333333} 32.3} & \cellcolor[HTML]{FFCCC9}{\color[HTML]{333333} 4.3} & \cellcolor[HTML]{34FF34}{\color[HTML]{333333} 74.1} & \cellcolor[HTML]{FD6864}{\color[HTML]{333333} 21.6} \\ \hline
\cellcolor[HTML]{FD6864}{\color[HTML]{333333} 37.6} & \cellcolor[HTML]{FD6864}{\color[HTML]{333333} 14.3} & \cellcolor[HTML]{9AFF99}{\color[HTML]{333333} 48.1} & \cellcolor[HTML]{FFFFFF}{\color[HTML]{333333} 0} & \cellcolor[HTML]{FD6864}{\color[HTML]{333333} 18.7} & \cellcolor[HTML]{34FF34}{\color[HTML]{333333} 81.3} & \cellcolor[HTML]{FFCCC9}{\color[HTML]{333333} 2.6} & \cellcolor[HTML]{FD6864}{\color[HTML]{333333} 16.4} & \cellcolor[HTML]{34FF34}{\color[HTML]{333333} 81} \\ \hline
\rowcolor[HTML]{C0C0C0} 
\multicolumn{9}{|c|}{\cellcolor[HTML]{9B9B9B}{\color[HTML]{333333} \textbf{M=100}}} \\ \hline
\rowcolor[HTML]{C0C0C0} 
\multicolumn{3}{|c|}{\cellcolor[HTML]{C0C0C0}\textbf{0 dB}} & \multicolumn{3}{c|}{\cellcolor[HTML]{C0C0C0}\textbf{5 dB}} & \multicolumn{3}{c|}{\cellcolor[HTML]{C0C0C0}\textbf{10 dB}} \\ \hline
\rowcolor[HTML]{EFEFEF} 
{\color[HTML]{333333} $\mathcal{H}_0$} & {\color[HTML]{333333} $\mathcal{H}_1$} & {\color[HTML]{333333} $\mathcal{H}_2$} & {\color[HTML]{333333} $\mathcal{H}_0$} & {\color[HTML]{333333} $\mathcal{H}_1$} & {\color[HTML]{333333} $\mathcal{H}_2$} & {\color[HTML]{333333} $\mathcal{H}_0$} & {\color[HTML]{333333} $\mathcal{H}_1$} & {\color[HTML]{333333} $\mathcal{H}_2$} \\ \hline
\cellcolor[HTML]{67FD9A}{\color[HTML]{333333} 58.2} & \cellcolor[HTML]{FFCCC9}{\color[HTML]{333333} 3} & \cellcolor[HTML]{FD6864}{\color[HTML]{333333} 38.8} & \cellcolor[HTML]{34FF34}{\color[HTML]{333333} 100} & {\color[HTML]{333333} 0} & {\color[HTML]{333333} 0} & \cellcolor[HTML]{34FF34}{\color[HTML]{333333} 100} & {\color[HTML]{333333} 0} & {\color[HTML]{333333} 0} \\ \hline
\rowcolor[HTML]{FFCCC9} 
{\color[HTML]{333333} 0.3} & \cellcolor[HTML]{34FF34}{\color[HTML]{333333} 90.6} & {\color[HTML]{333333} 9.1} & {\color[HTML]{333333} 0.7} & \cellcolor[HTML]{34FF34}{\color[HTML]{333333} 88.8} & \cellcolor[HTML]{FD6864}{\color[HTML]{333333} 10.5} & {\color[HTML]{333333} 1.1} & \cellcolor[HTML]{34FF34}{\color[HTML]{333333} 93.8} & {\color[HTML]{333333} 5.1} \\ \hline
\cellcolor[HTML]{FD6864}{\color[HTML]{333333} 37} & \cellcolor[HTML]{FFCCC9}{\color[HTML]{333333} 3.7} & \cellcolor[HTML]{67FD9A}{\color[HTML]{333333} 59.3} & {\color[HTML]{333333} 0} & \cellcolor[HTML]{FD6864}{\color[HTML]{333333} 10.7} & \cellcolor[HTML]{34FF34}{\color[HTML]{333333} 89.3} & \cellcolor[HTML]{FFCCC9}{\color[HTML]{333333} 0.2} & \cellcolor[HTML]{FFCCC9}{\color[HTML]{333333} 4.9} & \cellcolor[HTML]{34FF34}{\color[HTML]{333333} 94.9} \\ \hline
\end{tabular}}
\end{table*}

\par It is also observed from Tables \ref{title4} and \ref{title5} that the performance of the proposed jammer detection outperforms PUEA detection. This is because the jammer detection benefits from both the non-compressive nature of the jamming signal and the channel-dependent dictionary while the PUEA detection benefits only from the channel-dependent dictionary.

\par In classification, if a signal belongs class $i$ and is correctly classified in to belong to the same class, then it is said to be as true positive ($TP$). If it is wrongly classified to belong to a different class $j$, then it is said to be a false negative ($FN$). If, however, the signal does not belong to class $i$ and is wrongly classified as such, then it is counted as false positive ($FP$). Finally, if it does not really belong to $i$ and is classified to belong to $i$, then it is a true negative ($TN$). To this end, the true positive rate ($TPR$) or recall can be defined as $TPR={TP}/{(TP+FN)}$, whereas the false positive rate ($FPR$) can be defined as $FPR={FP}/{(FP+TN)}$. ROC curves and AUROC curve values show the capability of a classifier to distinguish between different classes. ROC is a probabilistic curve which is plotted with a $TPR$ on the vertical axis and $FPR$ on the horizontal axis. Ideally, the $TPR$ equals 1 and the $FPR$ equals 0. Generally speaking, the closer the ROC curve is to the top-left corner, the better the performance. Similarly, the higher values of the AUROC curve shows better performance. In this work, there are three classes ($\mathcal{H}_0$, $\mathcal{H}_1$, $\mathcal{H}_2$ or $\mathcal{H}_0$, $\mathcal{H}_1$, $\mathcal{H}_3$) and ROC curve for each class is plotted separately. 

\begin{figure}[b!]
\centering\resizebox{0.55\columnwidth}{!}{\includegraphics{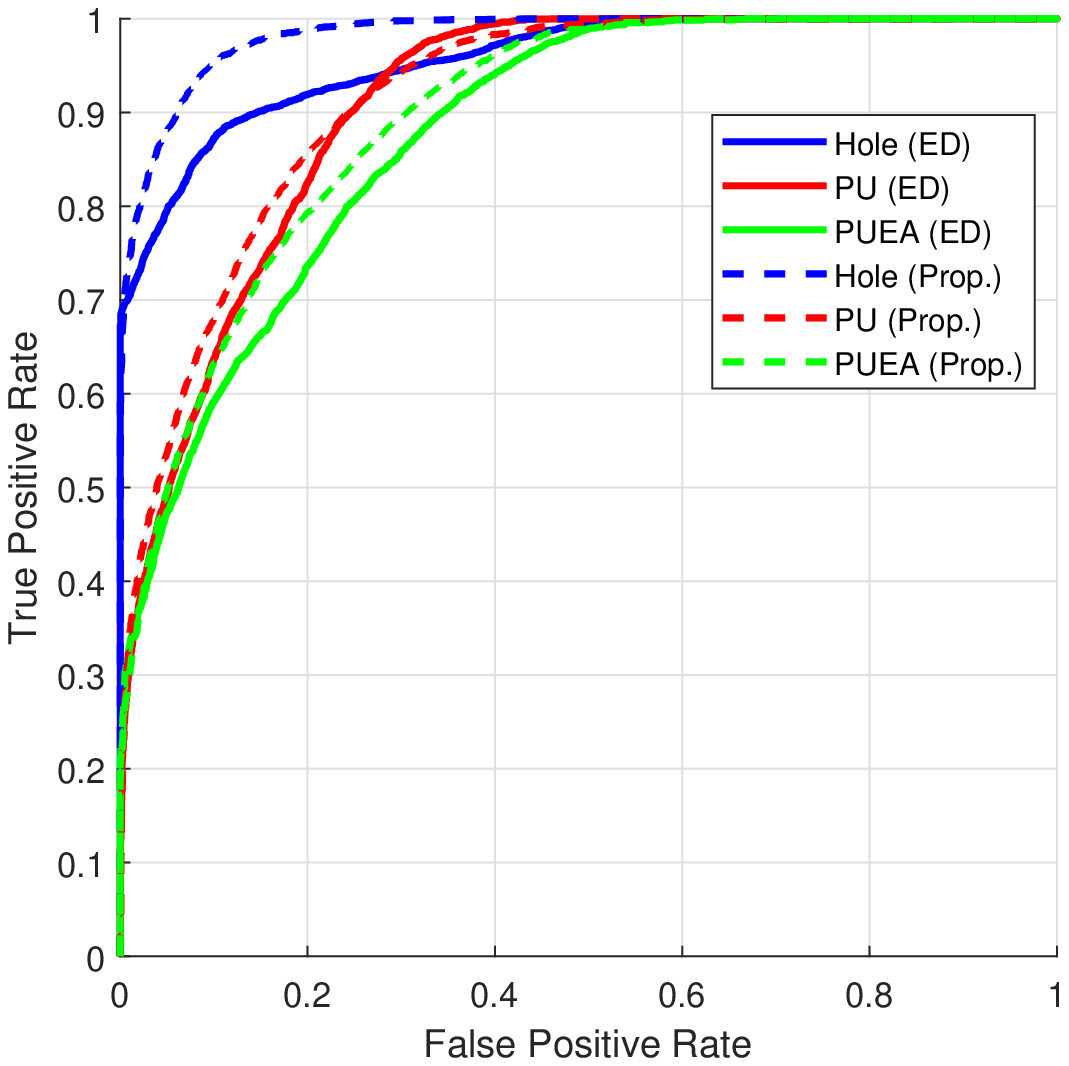}}
\caption{Comparison of the proposed algorithm with the ED-based ML algorithm for PUEA detection using ROC curves.}
\label{com1}
\end{figure}

\begin{figure}[t!]
\centering\resizebox{0.55\columnwidth}{!}{\includegraphics{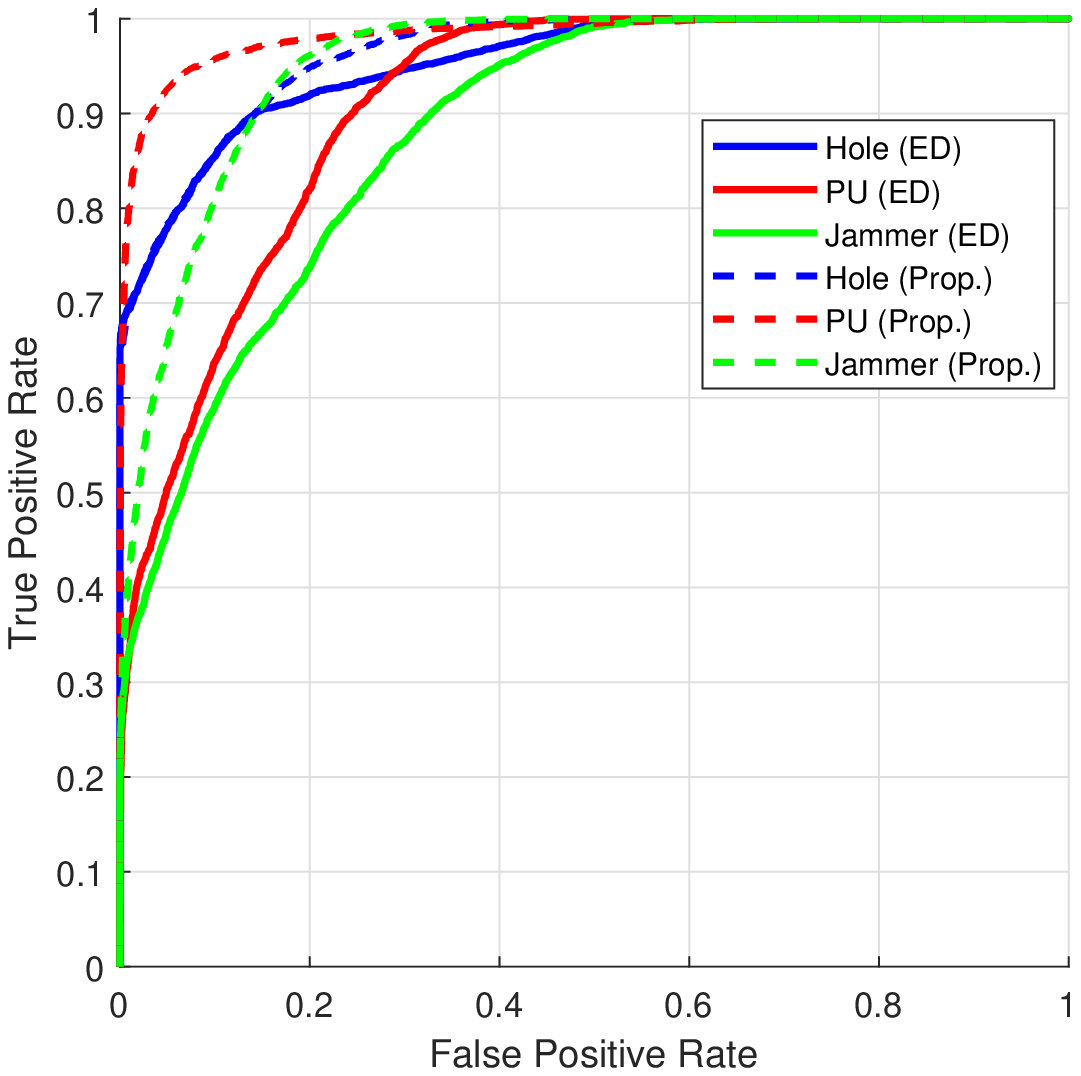}}
\caption{Comparison of the proposed algorithm with the ED-based ML algorithm for jamming attack detection using ROC curves.}
\label{com2}
\end{figure}

\par Fig.~\ref{com1} and Fig.~\ref{com2} present a performance comparison of the proposed algorithm with the ED-based ML algorithm for PUEA and jamming attack detection, respectively. In the case of ED-based ML, the energy of the received signals is used for the detection of different hypotheses while using ML structure similar to the one used for the proposed algorithm. It is observed from Fig.~\ref{com1} and Fig.~\ref{com2} that the ROC curves of the proposed algorithm are closer to the top-left corner compared to ROC curves of the ED-based algorithm for PUEA and jamming attack detection. Moreover, Tables \ref{aurocPUEA} and \ref{aurocjamming} also show that the values of AUROC of the proposed algorithms are higher compared to the AUROC values of the ED-based algorithms to detect different hypothesis presented in (\ref{equa8}). For example, the proposed algorithm outperforms the ED-based algorithm by 2.24 $\%$ in the case of PUEA and 6.88 $\%$ in case of jamming attack detection in terms of AUROC values. This is because the energy patterns in the residual and gradient vector enhance the detection capability of the proposed algorithm compared to the ED-based algorithm. 

\begin{table}[h!]
\caption {AUROC values for PUEA.} \label{aurocPUEA} 
\centering
\begin{tabular}{|c|c|c|c|}
\hline
 & \textbf{PU} & \textbf{Hole} & \textbf{PUE} \\ \hline
\textbf{ED-based} & 0.9089 & 0.9560 & 0.8719 \\ \hline
\textbf{Proposed} & 0.9152 & 0.9828 & 0.8943 
 \\ \hline
\end{tabular}
\end{table}

\begin{table}[h!]
\caption {AUROC values for jamming attack.} \label{aurocjamming} 
\centering
\begin{tabular}{|c|c|c|c|}
\hline
 & \textbf{PU} & \textbf{Hole} & \textbf{Jammer} \\ \hline
\textbf{ED-based} & 0.9097 & 0.9542 & 0.8820 \\ \hline
\textbf{Proposed} & 0.9830 & 0.9637 & 0.9508 \\ \hline
\end{tabular}
\end{table}

\par
\begin{figure}[h!]
\centering\resizebox{0.55\columnwidth}{!}{\includegraphics{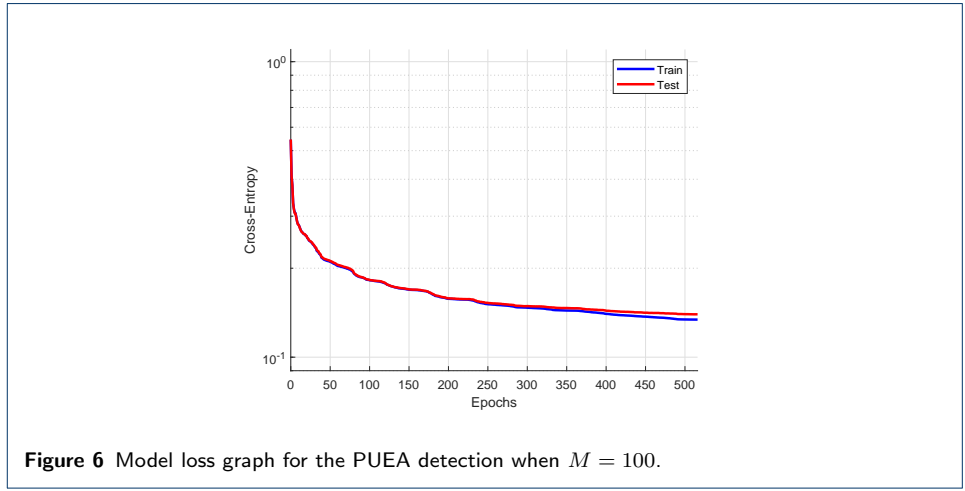}}
\caption{Model loss graph for the PUEA detection when $M=100$.}
\label{fig6}
\end{figure}

 From an ML point-of-view, a trained model (classifier in this work) should not memorize the inputs used in its training. To investigate this quality in the trained ML classifier model in the proposed algorithm, Fig.~\ref{fig6} shows the training and testing losses versus epochs for the PUEA detection when $M=100$. In view of this figure, it is evident that the accuracy of the training sets converges to the test set. These results signify the absence of overfitting, thereby validating the generalizability of the proposed model. In other words, the trained model does not memorize the training data. Here, it should be noted that we include the loss graph only for $M=100$ case of PUEA to avoid repetition. For the other values of $M$ and for the jammer case we observe the similar behavior in loss graphs.

\section{Conclusions}
\label{Section6}
\par In this paper, the convergence patterns of sparse recovery are exploited for the purpose of PUEA and jamming attack detection. Sparse recovery was conducted over a legitimate PU channel-dependent dictionary. Consequently, the signal from the legitimate node has smooth convergence as compared to the signal from the illegitimate node. Essentially, this awes to the fact that this signal is the only one compressible in the domain exclusively defined by this sparsifying dictionary. Besides, the non-compressive nature of a jamming signal with sparse coding over a PU channel-dependent dictionary was also exploited to detect jamming attacks. This detection algorithm made use of ML-based approaches. Numerical experiments showed the effectiveness of the proposed algorithm and its superior performance compared to ED-based ML algorithms. These results were validated in terms of confusion matrices, ROC curves, and values of AUROC curves, as quality metrics. In terms of AUROC curve values, the proposed algorithm outperformed the ED-based algorithm by 2.24 $\%$ in the case of PUEA and 6.88 $\%$ in case of jamming attack detection.

\appendix
\section*{Appendix Residual Energy Gradient Decay Analysis}
\par The proposed algorithm is based on the convergence patterns in the sparse coding of the compressed received signal. More specifically, the proposed algorithm uses a channel-dependent dictionary to identify different characteristics of gradients and residuals to detect PUEA and jamming attack.

\par In this work, we employ the computationally-efficient OMP for sparse coding. Let us focus on its first iteration for the sake of simplicity. At the start of the first OMP iteration, the signal itself is used to initialize the zero-$th$ residual $\boldsymbol{r}_0$. Afterwards, OMP chooses an atom ($\boldsymbol{d}$) from the atoms of the given dictionary $\boldsymbol{D_{PU}}$ that have the strongest similarity to the $\boldsymbol{r}_0$. This similarity is characterized by the projection corresponding to each atom as $\boldsymbol{E}= \boldsymbol{dd}^\dag$. The updated residual after the selection of atom can be given as
\begin{equation}
\boldsymbol{r}_1=\boldsymbol{r}_0 - \boldsymbol{E r}_0.
\label{equa16}
\end{equation}
\noindent For simplicity, the least-squares refinement of OMP is ignored. With each iteration, the residual magnitude is decreasing and the pattern of the concatenated residual values ($\|\boldsymbol{r}_1\|^{2}_{2}$) is used for classification. 

\par To this end, the first element in $\boldsymbol{G}$ can be represented as
\begin{equation}
\boldsymbol{G}(1)=\|\boldsymbol{r}_1\|^{2}_{2}-\|\boldsymbol{r}_0\|^{2}_{2}= \big \langle \boldsymbol{r_1},\boldsymbol{r_1}\big \rangle- 
\big \langle \boldsymbol{r_0},\boldsymbol {r_0}\big \rangle.
\label{equa17}
\end{equation}
\par Using this gradient magnitude property we can differentiate the cases $\mathcal{H}_0$, $\mathcal{H}_1$, $\mathcal{H}_2$, and $\mathcal{H}_3$. The general received signal can be given as: $\boldsymbol{y}=\boldsymbol{h}\boldsymbol{x}+\boldsymbol{n}$ and we can write the $\boldsymbol{G}(1)$ as follows
\begin{equation}
\boldsymbol{G}(1)=\|\boldsymbol{y-Ey}\|^{2}_{2}-\|\boldsymbol{y}\|^{2}_{2}.
\end{equation}

\par For the first hypothesis $\mathcal{H}_0$, $\boldsymbol{x}=0$. Thus, $\boldsymbol{y}$ is merely noise and can be written as
\begin{align}
\boldsymbol{G}(1)_{\mathcal{H}_0}&=\|\boldsymbol{n-En}\|^{2}_{2}-\|\boldsymbol{n}\|^{2}_{2} \nonumber ,\\
&=\big \langle\boldsymbol{n}-\boldsymbol{En},\boldsymbol{n}-\boldsymbol{En}\big \rangle-\big \langle\boldsymbol{n},\boldsymbol{n}\big \rangle
\nonumber ,\\
&= \big \langle\boldsymbol{n},\boldsymbol{n}\big \rangle-2\big \langle\boldsymbol{n},\boldsymbol{En}\big \rangle+\big \langle\boldsymbol{En},\boldsymbol{En}\big \rangle-\big \langle\boldsymbol{n},\boldsymbol{n}\big \rangle.
\label{equa18a}
\end{align}
\par With respect to the properties of projection, we know that $\big \langle\boldsymbol{En},\boldsymbol{n}\big \rangle = \big \langle\boldsymbol{En},\boldsymbol{En}\big \rangle = \|\boldsymbol{En}_2^{2}\|$.
Hence, (\ref{equa18a}) can be written as
\begin{equation}
\boldsymbol{G}(1)_{\mathcal{H}_0}=-\|\boldsymbol{En}\|_2^{2}.
\label{equa18aa}
\end{equation}

\par Following the same logic, $\boldsymbol{G}(1)$ for $\mathcal{H}_1$, $\mathcal{H}_2$ and $\mathcal{H}_3$ can be expressed as follows
 \begin{align}
 \boldsymbol{G}(1)&=\big \langle\boldsymbol{hx+n}-\boldsymbol{E(hx+n)}, \boldsymbol{hx}+\boldsymbol{n}-\boldsymbol{E(hx+n)}\big \rangle\nonumber\\&\qquad\displaystyle-\big \langle\boldsymbol{hx}+\boldsymbol{n},\boldsymbol{hx}+\boldsymbol{n}\big \rangle, \nonumber \\
 &=\boldsymbol{a}-2\boldsymbol{b}+\boldsymbol{c}-\boldsymbol{d} \label{qwer},
 \end{align}
\noindent where $\boldsymbol{a}$, $\boldsymbol{b}$, $\boldsymbol{c}$ and $\boldsymbol{d}$ are defined next. Specifically, $\boldsymbol{a}$ can be written as
 \begin{align}
 \boldsymbol{a}&=\big \langle\boldsymbol{h}\boldsymbol{x}+\boldsymbol{n},\boldsymbol{h}\boldsymbol{x}+\boldsymbol{n}\big \rangle, \nonumber \\
 &=\big \langle\boldsymbol{h}\boldsymbol{x}+\boldsymbol{h}\boldsymbol{x}\big \rangle+\big \langle\boldsymbol{h}\boldsymbol{x}+\boldsymbol{n}\big \rangle+\big \langle\boldsymbol{h}\boldsymbol{x}+\boldsymbol{n}\big \rangle+\big \langle\boldsymbol{n}+\boldsymbol{n}\big \rangle, \nonumber\\
&=\big \langle\boldsymbol{h}\boldsymbol{x}+\boldsymbol{h}\boldsymbol{x}\big \rangle+2\big \langle\boldsymbol{h}\boldsymbol{x}+\boldsymbol{n}\big \rangle+\big \langle\boldsymbol{n}+\boldsymbol{n}\big \rangle. \label{gre}
\end{align}

\par Assuming that the noise is independent of $\boldsymbol{h}\boldsymbol{x}$, $\big \langle\boldsymbol{h}\boldsymbol{x}+\boldsymbol{n}\big \rangle=0$, 
we can write (\ref{gre}) as
\begin{align}
\boldsymbol{a}=\big \langle\boldsymbol{h}\boldsymbol{x}+\boldsymbol{h}\boldsymbol{x}\big \rangle+\big \langle\boldsymbol{n}+\boldsymbol{n}\big \rangle.
\label{aval}
\end{align}

\par By its turn, $\boldsymbol{b}$ can be expressed as
\begin{align}
\boldsymbol{b}&=\big \langle \boldsymbol{hx+n}, \boldsymbol{E(hx+n)}\big \rangle ,\nonumber \\ 
&=\big \langle\boldsymbol{hx+n},\boldsymbol{Ehx+En}\big \rangle, \nonumber \\ 
&=\big \langle\boldsymbol{h}\boldsymbol{x},\boldsymbol{Eh}\boldsymbol{x}\big \rangle+\big \langle\boldsymbol{h}\boldsymbol{x},\boldsymbol{En}\big \rangle+\big \langle\boldsymbol{n},\boldsymbol{Eh}\boldsymbol{x}\big \rangle+\big \langle\boldsymbol{n},\boldsymbol{En}\big \rangle, \nonumber \\
&= \big \langle\boldsymbol{Eh}\boldsymbol{x},\boldsymbol{Eh}\boldsymbol{x}\big \rangle+\big \langle\boldsymbol{h}\boldsymbol{x},\boldsymbol{En}\big \rangle+\big \langle\boldsymbol{n},\boldsymbol{Eh}\boldsymbol{x}\big \rangle+\big \langle\boldsymbol{En},\boldsymbol{En}\big \rangle, \nonumber \\
&= \big \langle\boldsymbol{Eh}\boldsymbol{x},\boldsymbol{Eh}\boldsymbol{x}\big \rangle+\big \langle\boldsymbol{En},\boldsymbol{En}\big \rangle, \label{bval}
\end{align}
\noindent where $\big \langle\boldsymbol{hx},\boldsymbol{En}\big \rangle=\big \langle\boldsymbol{n},\boldsymbol{Eh}\boldsymbol{x}\big \rangle=0$.

\par Moreover, $\boldsymbol{c}$ can be expressed as
\begin{align}
\boldsymbol{c}&=\big \langle\boldsymbol{E}(\boldsymbol{\boldsymbol{h}\boldsymbol{x}+\boldsymbol{n}}),\boldsymbol{E}(\boldsymbol{\boldsymbol{h}\boldsymbol{x}+\boldsymbol{n}})\big \rangle, \nonumber \\ 
&=\big \langle\boldsymbol{Eh}\boldsymbol{x}+\boldsymbol{En},\boldsymbol{Eh}\boldsymbol{x}+\boldsymbol{En}\big \rangle \nonumber,\\
&=\big \langle\boldsymbol{Eh}\boldsymbol{x},\boldsymbol{Ehx}\big \rangle+\big \langle\boldsymbol{En},\boldsymbol{En}\big \rangle.
\label{cval}
\end{align}

\par Lastly, $\boldsymbol{d}$ can be given as
\begin{align}
\boldsymbol{d}=\big \langle\boldsymbol{h}\boldsymbol{x}+\boldsymbol{h}\boldsymbol{x}\big \rangle+\big \langle\boldsymbol{n}+\boldsymbol{n}\big \rangle=\boldsymbol{a}.
\label{dval}
\end{align}

\par Based on (\ref{aval}), (\ref{bval}), (\ref{cval}) and (\ref{dval}), and making the appropriate substitution, $\boldsymbol{G}(1)$ can be written as
\begin{align}
\boldsymbol{G}(1)&=\boldsymbol{a}-2\boldsymbol{b}+\boldsymbol{c}-\boldsymbol{d}, \nonumber \\
&=-2\big \langle\boldsymbol{Eh}\boldsymbol{x},\boldsymbol{Eh}\boldsymbol{x}\big \rangle-2\big \langle\boldsymbol{En},\boldsymbol{En}\big \rangle\nonumber\\&\qquad+\big \langle\boldsymbol{Eh}\boldsymbol{x},\boldsymbol{Eh}\boldsymbol{x}\big \rangle+\big \langle\boldsymbol{En},\boldsymbol{En}\big \rangle, \nonumber \\
&=-\big \langle\boldsymbol{Ehx},\boldsymbol{Ehx}\big \rangle-\big \langle\boldsymbol{En},\boldsymbol{En}\big \rangle .
\end{align}

\par Finally, the generic expression of the gradient magnitude for hypotheses $\mathcal{H}_1$, $\mathcal{H}_2$ and $\mathcal{H}_3$ can be expressed
\begin{equation}
\boldsymbol{G}(1)_{\mathcal{H}_1,\mathcal{H}_2,\mathcal{H}_3}=-\|\boldsymbol{Ehx}\|_2-\|\boldsymbol{En}\|_2,\label{20}
\end{equation} 
\noindent where $\boldsymbol{h}$ corresponds to $\boldsymbol{h}_{PU}$ in case of PU or $\boldsymbol{h}_{i}$ in case of PUE/jammer as explained in (\ref{equa8}). Moreover, $\boldsymbol{x}$ will be structured in case of PU and PUE, while unstructured in the case of a jamming attack.

\begin{backmatter}
\section*{Abbreviations}
AUROC: Area under receiver operating characteristics; CR: Cognitive radio; CS: Compressive sensing; ED: Energy detection; FFT: Fast Fourier transform; $FN$: False-negative; $FP$: False positive; $FPR$: False positive rate; FSK: Frequency-shift keying; OMP: Orthogonal matching pursuit; PSK: Phase-shift keying; PAM: Pulse amplitude modulation; PU: Primary user; PUE: Primary user emulator; PUEA: Primary user emulation attack; QAM: Quadrature amplitude modulation; QoS: Quality of service; ML: Machine learning; ROC: Receiver operating characteristics; SNR: Signal-to-noise ratio; SSDF: Spectrum sensing data falsification; SU: Secondary Users; SVDD: Support vector data description; $TN$: True negative; $TPR$: True positive rate; $TP$: True positive. 
\section*{Availability of data and materials}
Please contact the corresponding author at haji.madni@std.medipol.edu.tr.
\section*{Author's contributions}
HM, MA, and MN developed the main idea under the supervision of H. A. \footnote{H. Arslan was supported by the Scientific and Technological Research Council of Turkey (TUBITAK) under Grant 119E433.} They implemented the basic simulations and drafted the manuscript jointly. All authors participated in the development of the main idea, its implementation, and the interpretation of the results. All authors have read and approved the manuscript.
\section*{Competing interests}
The authors declare that they have no competing interests.
%\section*{Acknowledgment}
%The work of H. Arslan was supported by the Scientific and Technological Research Council of Turkey (TUBITAK) under Grant 119E433.
%The authors are thankful to M. Sohaib J. Solaija and Farwa Ahmed for their suggestions and comments. 

%\bibliographystyle{bmc-mathphys} 

\end{backmatter}
\section*{Figures Title and Legend Section}
\textbf{Figure 1: Title}$\{$The basic system model$\}$\\
\textbf{Figure 1: Legend}$\{$The basic system model, where PUE and jammer want to degrade SU's spectrum utilization by sending a fake signal.$\}$\\
\textbf{Figure 2: Title}$\{$Experimental motivation$\}$\\
\textbf{Figure 2: Legend}$\{$The averages of $\|\boldsymbol{r}\|_2$ versus sparse coding iteration for received signals under hypotheses $\mathcal{H}_0$, $\mathcal{H}_1$, $\mathcal{H}_2$ and $\mathcal{H}_3$ are presented in sub-figures (a), (b), (c), and (d), respectively, while the averages of $|\boldsymbol{G}|$ versus sparse coding iteration are presented in sub-figures (e), (f), (g), and (h), respectively.$\}$\\
\textbf{Figure 3: Title}$\{$The proposed algorithm$\}$\\
\textbf{Figure 3: Legend}$\{$An illustration of the proposed algorithm (a) training stage, (b) testing stage.$\}$\\
\textbf{Figure 4: Title}$\{$ROC curves for PUEA detection performance analysis$\}$\\
\textbf{Figure 4: Legend}$\{$Comparison of the proposed algorithm with the ED-based ML algorithm for PUEA detection using ROC curves.$\}$\\
\textbf{Figure 5: Title}$\{$ROC curves for jamming attack detection performance analysis$\}$\\
\textbf{Figure 5: Legend}$\{$Comparison of the proposed algorithm with the ED-based ML algorithm for jamming attack detection using ROC curves.$\}$\\
\textbf{Figure 6: Title}$\{$Model loss graph$\}$\\
\textbf{Figure 6: Legend}$\{$Model loss graph for the PUEA detection when $M=100$$\}$\\
\end{document}